\documentclass{llncs}
\usepackage{endnotes}

\usepackage[utf8]{inputenc}
\usepackage[T1]{fontenc}

\usepackage[table]{xcolor}
\usepackage{graphicx}
\usepackage{caption}
\usepackage{subcaption}
\usepackage{stfloats}
\usepackage{float}
\usepackage[weather]{ifsym}

\usepackage[hyphens]{url}
\usepackage[colorlinks=true, urlcolor=black, linkcolor=black]{hyperref}
\usepackage{paralist}
\renewenvironment{itemize}{\begin{compactitem}}{\end{compactitem}}
\renewenvironment{enumerate}{\begin{compactenum}}{\end{compactenum}}

\setdefaultenum{(1.)}{}{}{}
\setdefaultitem{$\blacktriangleright$}{}{}{}
\usepackage{listings}
\newcommand{\todo}[1]{{\color{red}#1}}
\newcommand{\evalnumber}[1]{#1}

\renewcommand*{\paragraph}[1]{\vspace{2mm}\noindent\textbf{#1.}}
\usepackage{setspace}
\usepackage{array}
\newcolumntype{x}[1]{>{\centering\let\newline\\\arraybackslash\hspace{0pt}}p{#1}}
\usepackage{threeparttable}

\usepackage{amsmath,amssymb,wasysym}
\DeclareMathAlphabet{\mathcal}{OMS}{cmsy}{m}{n}
\newcommand{\A}{\ensuremath{\mathcal{A}}}

\newcommand{\C}{\ensuremath{\mathcal{C}}}

\newcommand{\I}{\ensuremath{\mathcal{I}}}

\newcommand{\V}{\ensuremath{\mathcal{V}}}

\usepackage{comment}
\specialcomment{overview}{\textbf{Overview:}}{}
\excludecomment{overview}

\newcommand{\IdP}{\ensuremath{\I{}dP}}
\newcommand{\IdPA}{\ensuremath{\IdP{}_\A}}
\newcommand{\IdPV}{\ensuremath{\IdP{}_\V}}

\newcommand{\ID}{\ensuremath{\text{ID}}}
\newcommand{\URLID}{\ensuremath{\text{URL.ID}}}
\newcommand{\URLIdP}{\ensuremath{\text{URL.}\IdP}}
\newcommand{\URLSP}{\ensuremath{\text{URL.}SP}}
\newcommand{\URLA}{\ensuremath{\text{URL.}\A}}

\usepackage[xindy]{glossaries}

\glossarystyle{indexgroup}
\makeglossaries

\newacronym[]{trc}{TRC}{Token Recipient Confusion}
\newacronym[]{kc}{KC}{Key Confusion}
\newacronym[]{ids}{IDS}{ID Spoofing}
\newacronym[]{ds}{DS}{Discovery Spoofing}
\newacronym[%
description={\glsentrylong{aaas}; A SOA concept which provides an authentication service as a service},
]
{aaas}{AaaS}{authentication as a service}
\newacronym[%
description={\glsentrylong{ax}; OpenID extension for exchanging identity information between endpoints~\cite{ax}},
]
{ax}{Ax}{Attribute Exchange}
\newacronym[%
    description={\glsentrylong{sreg}; OpenID extension for exchanging identity information between endpoints~\cite{sreg}},
]
{sreg}{SReg}{Simple Registration}
\newacronym[%
description={\glsentrylong{cms}; A software which helps to create an manage content collaboratory. CMSs are often web applications},
]
{cms}{CMS}{content-management system}
\newacronym[%
description={\glsentrylong{idp}; An entity which is responsible for creating tokens that will be used to authenticate a user to an SP},
]
{idp}{IdP}{Identity Provider}
\newacronym[%
description={\glsentrylong{dos}; The unavailability of a service, which should be available},
]
{dos}{DoS}{Denial-of-Service}
\newacronym[%
description={\glsentrylong{gui}; A software component which allows a human to interact with machine by using symbols},
]
{gui}{GUI}{Graphical User Interface}
\newacronym[%
description={\glsentrylong{jaxb}; An interface for Java allowing to bind XML Data to Java objects. See \url{https://jaxb.java.net/}},
]
{jaxb}{JAXB}{Java Architecture for XML Binding}
\newacronym[%
description={\glsentrylong{mitm}; An attack technique in which the attacker virtually or physically sits between to users, allowed to listen and manipulate their exchanged messages},
]
{mitm}{MitM}{Man-in-the-Middle}
\newacronym[
    description={Single Sign-On is a property of access controll, which allows a user to log in once and request access to several systems},
]{sso}{SSO}{Single Sign-On}
\newacronym[%
description={\glsentrylong{soa}; Abstract model of software architecture},
]
{soa}{SOA}{Service Oriented Architecture}
\newacronym[%
description={\glsentrylong{sp}; Also known as \emph{Relying Party}, entity that offers a service which the user wants to use},
]
{sp}{SP}{Service Provider}
\newacronym[%
description={A user agent is software (a software agent) that is acting on behalf of a user (Wikipedia)}
]
{ua}{UA}{user agent}
\newacronym[%
description={\glsentrylong{w3c}; Organization for standards in the World Wide Web},
]
{w3c}{W3C}{World Wide Web Consortium}
\newacronym[%
description={\glsentrylong{xml}; Textual data format to encode documents, commonly used for message exchange~\cite{xml}},
]
{xml}{XML}{eXtended Markup Language}
\newacronym[%
description={\glsentrylong{xsw}; Technique for attacking signed XML documents~\cite{McIntosh2005}},
]
{xsw}{XSW}{XML Signature Wrapping}

\newglossaryentry{dhke}{%
    name={Diffie-Hellman key exchange},
    description={Specific method for exchanging key material},
}
\newglossaryentry{asso}{%
    name={association},
    description={An association between the SP and the OpenID IdP establishes a shared secret between them~\cite[Section 8]{openid20}},
}
\newglossaryentry{csrf}{%
    name={CSRF},
    description={Cross-Site-Request-Forgery; Attack technique which tries to use web application APIs by a victim without his knowledge},
}
\newglossaryentry{dirver}{%
    name={direct verification},
    description={OpenID signature verfication performed by the OpenID IdP itself, enforced by the SP~\cite[Section 11.4.2]{openid20}},
}
\newglossaryentry{junit}{%
    name={JUnit},
    description={JUnit is framework for testing Java programs},
}
\newglossaryentry{testng}{%
    name={TestNG},
    description={TestNG is framework for testing Java programs},
}
\newglossaryentry{mysql}{%
    name={MySQL},
    description={MySQL is a widely used \glsentrytext{opensource} relational database management system},
}
\newglossaryentry{oauth}{%
    name={OAuth},
    description={OAuth is an open protocol which standardized API-Authoring for desktop-, web-, and mobile-applications~\cite{oauth}},
}
\newglossaryentry{oid}{%
    name={OpenID},
    description={OpenID is decentralized authentication system for web-based SPs~\cite{openid20}},
}
\newglossaryentry{oid4j}{%
    name={OpenID4Java},
    description={OpenID4Java is library which easily allows to implement an OpenID consumber as well as an OpenID IdP ~\cite{openid4java}},
}
\newglossaryentry{oida}{%
    name={OpenID Attacker},
    description={OpenID Attacker is the tool developed in this thesis for attacking the OpenID specification},
}
\newglossaryentry{opensource}{%
    name={open source},
    description={Work which are enforced by license to have source available for the public},
}
\newglossaryentry{oc}{
    name={ownCloud},
    description={OwnCloud is a free and \glsentrytext{opensource}, \glsentrytext{php} and \glsentrytext{mysql} based web application which allows data synchronization and cloud storage~\cite{owncloud}.}
}
\newglossaryentry{pentest}{%
    name={penetration test},
    user1={Penetration Test},
    user2={Penetration Tests},
    user3={penetration testing},
    user4={Penetration Testing},
    user5={penetration tester},
    user6={penetration testers},
    description={Method for evaluating security on computer systems},
}
\newglossaryentry{php}{%
    name={PHP},
    description={PHP is a popular server-side scripting language and widespread in the area of web development},
}
\newglossaryentry{saml}{%
    name={SAML},
    description={Security Assertion Markup Language~\cite{saml}},
}
\newglossaryentry{sqlinjection}{
    name={SQL-Injection},
    description={An attack technique which tries to inject and execute SQL statements reasoned in inadequate user input validation~\cite{sqlinjection}},
}
\newglossaryentry{wsattacker}{%
    name={WS-Attacker},
    description={Automatic \glsentrytext{pentest} framework~\cite{wsattacker}},
}
\newglossaryentry{doc}{%
    name={XML document},
    description={General term for a document which uses \glsentrytext{xml} as description language, e.g.\ a \glsentrytext{soap} message},
    parent=xml,
    user1={document},
    user2={documents},
}
\newglossaryentry{wa}{%
    name={web application},
    user1={Web Application},
    user2={Web Applications},
    description={A web application is an application that uses a web browser as a client.}
}
\newglossaryentry{wp}{%
    name={WordPress},
    description={WordPress is a free and \glsentrytext{opensource}, \glsentrytext{php} and \glsentrytext{mysql} based \glsentrytext{cms}~\cite{wordpress}.}
}
\newglossaryentry{xmlsig}{%
    name={XML Signature},
    user1={XML Digital Signature},
    description={also \glsentryuseri{xmlsig}; Standard for creating signatures in \glsentrytext{xml} documents~\cite{xmlsig1, xmlsig2}},
}
\newglossaryentry{xrds}{%
    name={XRDS},
    description={eXtensible Resource Descriptor Sequence is an XML format for describing metadata as a web resource},
}
\newglossaryentry{xss}{%
    name={XSS},
    description={Cross-Site-Scripting is a web application vulnerability which tries to execute an attacker script within the context of the website owner within the user's browser},
}

\author{Christian Mainka\thanks{This work has been funded by the European Union within the European Regional Development Fund program.} \and Vladislav Mladenov\thanks{The author was supported by the SkIDentity project of the German Federal Ministry of Economics and Technology (BMWi, FKZ: 01MD11030).} \and J\"org Schwenk}
\authorrunning{Mainka et al.} % abbreviated author list (for running head)
\tocauthor{Christian Mainka, Vladislav Mladenov, J\"org Schwenk}
\institute{
Horst G\"ortz Institute for IT Security, Ruhr-University Bochum, Germany\\
\email{\{christian.mainka,vladislav.mladenov,joerg.schwenk\}@hgi.rub.de}
}

\begin{document}

\title{Do not trust me: Using malicious IdPs for analyzing and attacking Single Sign-On}
\maketitle
\begin{abstract}
\gls{sso} systems simplify login procedures by using an an \gls{idp} to issue authentication tokens which can be consumed by \glspl{sp}. Traditionally, \glspl{idp} are modeled as trusted third parties. This is reasonable for SSO systems like Kerberos, MS Passport and SAML, where each \gls{sp} explicitely specifies which \gls{idp} he trusts. However, in open systems like OpenID and OpenID Connect, each user may set up his own \gls{idp}, and a discovery phase is added to the protocol flow. Thus it is easy for an attacker to set up its own \gls{idp}.

In this paper we use a novel approach for analyzing \gls{sso} authentication schemes by introducing a {\em malicious \gls{idp}}. With this approach we evaluate one of the most popular and widely deployed \gls{sso} protocols -- \gls{oid}. 
We found four novel attack classes on \gls{oid}, which were not covered by previous research, and show their applicability to real-life implementations. As a result, we were able to compromise \evalnumber{11} out of \evalnumber{16} existing \gls{oid} implementations like Sourceforge, Drupal and \gls{oc}.

We automated discovery of these attacks in a \gls{opensource} tool {\em \gls{oida}}, which additionally allows fine-granular testing of all parameters in \gls{oid} implementations.

Our research helps to better understand the message flow in the \gls{oid} protocol, trust assumptions in the different components of the system, and implementation issues in \gls{oid} components. It is applicable to other \gls{sso} systems like OpenID Connect and SAML.
% We propose generic fixes. 
All \gls{oid} implementations have been informed about their vulnerabilities and we supported them in fixing the issues.

\end{abstract}

% \category{H.4}{Information Systems Applications}{Miscellaneous}
%A category including the fourth, optional field follows...
% \category{D.2.8}{Software Engineering}{Metrics}[complexity measures, performance measures]

% \terms{Theory}

% \keywords{\gls{sso}, \gls{oid}, impersonation attacks}

% \input{todo}

\section{Introduction}

\newcommand{\SESSION}{\texttt{\$\_SESSION}}
\colorlet{myred}{red!60!black}
% \colorlet{mygreen}{green!40!black}
\colorlet{mygreen}{black}
\begin{table*}[t]
    \centering
    \newcounter{targets}
    \newcounter{RC}
    \newcounter{IDS}
    \newcounter{DS}
    \newcounter{KC}
    \newcounter{theft}
    \newcommand{\inred}[1]{\textcolor{myred}{#1}}
    \newcommand{\ingreen}[1]{\textcolor{mygreen}{#1}}
    \newcommand{\symbolyes}{\inred{\Lightning}}
    \newcommand{\yes}[1]{\symbolyes\addtocounter{#1}{1}}
    \newcommand{\symbolYES}{\inred{\Lightning\Lightning}}
    \newcommand{\YES}[1]{\symbolYES\addtocounter{#1}{1}}
    \newcommand{\symbolno}{\ingreen{-}}
    \newcommand{\no}{\symbolno}

{\small
    \rowcolors{2}{gray!10}{}
    \begin{tabular}{|p{0.368\linewidth}x{0.15\linewidth}*{4}{x{0.07\linewidth}}|x{0.13\linewidth}|}
            \hline
            \textbf{Service Provider} & \textbf{Programming Language} & \textbf{\acrshort{trc}}  & \textbf{\acrshort{kc}}& \textbf{\acrshort{ids}} & \textbf{\acrshort{ds}} & \textbf{Summary: Unauthorized Access}  \\
            \hline                                                                                                                     
            CF OpenID                                         & ColdFusion              & \yes{RC} & \no      & \YES{IDS} & \no      & \YES{theft} \addtocounter{targets}{1}\\
            DotNet OpenAuth                                   & \.NET                   & \no      & \no      & \no       & \no      & \no  \addtocounter{targets}{1}\\
            Drupal 6 / Drupal 7                               & \gls{php}               & \no      & \YES{KC} & \no       & \no      & \YES{theft} \addtocounter{targets}{1}\\
            dyuproject                                        & Java                    & \yes{RC} & \no      & \YES{IDS} & \no      & \YES{theft} \addtocounter{targets}{1}\\
            janrain                                           & \gls{php}, Python, Ruby & \no      & \no      & \no       & \no      & \no  \addtocounter{targets}{1}\\
            JOID                                              & Java                    & \yes{RC} & \no      & \YES{IDS} & \no      & \YES{theft} \addtocounter{targets}{1}\\
            JOpenID                                           & Java                    & \no      & \no      & \YES{IDS} & \no      & \YES{theft} \addtocounter{targets}{1}\\
            libopkele \newline (Apache mod\_auth\_openid)     & C++                     & \no      & \no      & \no       & \no      & \no  \addtocounter{targets}{1}\\
            LightOpenID                                       & \gls{php}               & \no      & \no      & \no       & \no      & \no  \addtocounter{targets}{1}\\
            Net::OpenID::Consumer                             & Perl                    & \yes{RC} & \no      & \no       & \no      & \yes{theft} \addtocounter{targets}{1}\\
            OpenID 4 Java (WSO2)                              & Java                    & \no      & \no      & \no       & \no      & \no  \addtocounter{targets}{1}\\
            OpenID CFC                                        & ColdFusion              & \no      & \no      & \YES{IDS} & \no      & \YES{theft} \addtocounter{targets}{1}\\
            OpenID for Node.js \newline (everyauth, Passport) & JavaScript/NodeJS      & \yes{RC}  & \no      & \no       & \no      & \yes{theft} \addtocounter{targets}{1}\\
            Simple OpenID PHP Class \newline (\gls{oc} 5)     & \gls{php}               & \yes{RC} & \no      & \no       & \YES{DS} & \yes{theft} \addtocounter{targets}{1}\\
            Sourceforge                                       & n.a.                    & \no      & \YES{KC} & \YES{IDS} & \no      & \YES{theft} \addtocounter{targets}{1}\\
            Zend Framework \newline OpenID Component          & \gls{php}               & \no      & \YES{KC} & \no       & \no      & \YES{theft} \addtocounter{targets}{1}\\
            \hline
            \hline
            Total                                             &                         & \theRC   & \theKC   & \theIDS   & \theDS   & \thetheft / \thetargets \\
                                                               % &  & \theRC & \theKC & \theIDS & \theDS & 8 $\times$ \symbolYES \\
                                                               % &  &        &        &         &        & 3 $\times$ \symbolyes \\
                                                         % Total &  &        &        &         &        & \thetheft / \thetargets \\
            \hline
    \end{tabular}
}
\vspace{1mm}
{\small
    \begin{tabular}{ll}
        \symbolyes & One account on the target is compromised. \\
        \symbolYES & All accounts on the target are compromised. \\
    \end{tabular}
}
  \vspace{2mm}
  \caption{Results of our practical evaluation. For eleven out of 16 targets, we could get \emph{unauthorized access}. Three targets were compromised using the web attacker model (\textcolor{myred}{\Lightning}). The other eight targets make use of a weaker variant (\textcolor{myred}{\Lightning\Lightning}), \emph{without any user interaction}.}
  \label{tab:results}
\end{table*}

\paragraph{Single Sign-On}
\Glsfirst{sso} is a technique to enhance and simplify the login process on websites.
Instead of managing a plethora of username/password combinations for each website, a user just needs an account at an \glsfirst{idp} which can then be used to log in on a \glsfirst{sp}.

\begin{figure}[!ht]
    \centering
    \includegraphics[width=1.0\linewidth]{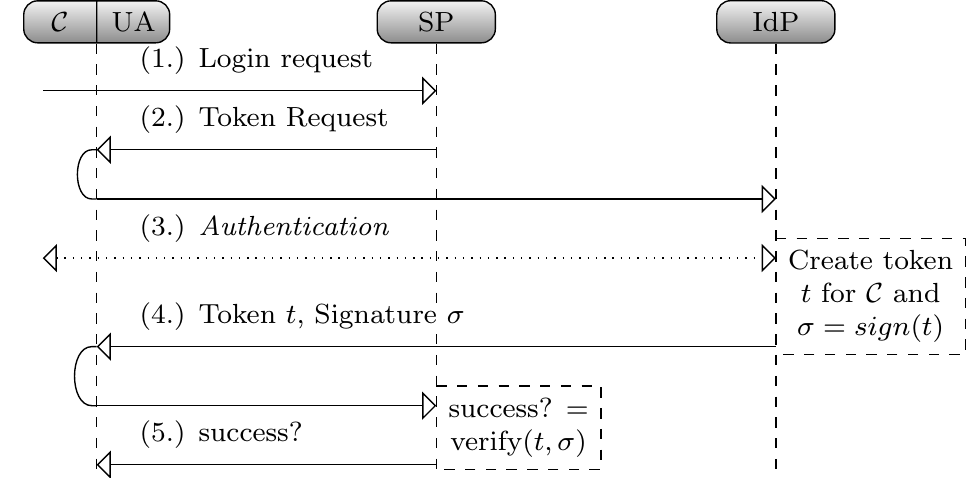}
    \caption{\glsfirst{sso} overview.}
    \label{fig:sso_overview}
\end{figure}

\autoref{fig:sso_overview} gives an overview of a basic \gls{sso} scenario.
When a client ($\C$) tries to log in to a service offered by the \gls{sp}, $\C$ sends a login request (1.) through some \gls{ua} (typically a webbrowser).
If $\C$ is not yet authenticated to the \gls{sp}, a token request is returned (2.).
The token request message contains information on the \gls{sp}, the chosen \gls{idp} (e.g.\ the \gls{idp}'s URL) and optionally on $\C$'s account name at the \gls{idp}.
$\C$'s user agent is redirected to the \gls{idp} and forwards the token request to it.
If $\C$ is not yet logged in at this \gls{idp}, she/he has to authenticate in Step (3.).
The \gls{idp} then issues an authentication token $t$ for $\C$ which is commonly protected by a cryptographic signature\footnote{In many specifications, both, Message Authentication Codes (MACs) and Digital Signatures, are summarized under the term \emph{signature}.} $\sigma$.
In Step (4.), $t$ is sent back to the \gls{ua}, which forwards it to the \gls{sp}.
Finally, the \gls{sp} verifies $t$ and, in case of successful verification, grants access to its resources in Step (5.).

\paragraph{Motivation} 
Password based authentication still dominates the Internet, but security problems related to passwords are obvious: Users either use weak passwords or reuse passwords between different sites, password based login is prone to simple  attacks like Phishing or dictionary based attacks, and recently two studies on password managers \cite{Silver2014,Li2014a} showed all of them to be insecure.
\gls{sso} schemes have been proposed to replace password based authentication, to enhance both usability and security. A recent non-academic overview~\cite{blueresearch} claims that 87\% of U.S.\ customers are aware of \gls{sso} and more than half have tried it. \gls{oid} is one of the most widespread \gls{sso} protocols and is currently integrated in 1.2 million websites~\cite{openidstatistics}. Leading companies like Google, Facebook, and PayPal support \gls{oid} based authentication.

The prospect of enhanced security through the introduction of \gls{sso} schemes is combined with higher risks because \gls{sso} schemes constitue a {\em single point of attack}:
If a weakness in a SSO scheme is detected, a large number of \acrlong{sp}s on the Internet may be affected simultaneously. 
Thus from the beginning, SSO schemes have been subject to formal security analysis~\cite{grossSAML,GP06SAML}.
Wang et al.~\cite{microsoft} initiated a new branch of research on SSO protocols by analyzing messages exchanged in (partly undocumented) real-life implementations, which led to the identification of serious logic flaws. 
They introduced a tool called BRM Analyzer to assist in the analysis of implementations.
Their analysis only considered messages that could be seen by the browser, and omitted the information flow between \gls{sp} and \gls{idp}. 
In their model (and in all other previous work, cf. \autoref{sec:relatedwork}), client and \gls{sp} may be controlled by the attacker, but the \gls{idp} is assumed to be trustworthy.

In view of the importance of \gls{sso} and \gls{oid}, and of the impact a single vulnerability in a SSO system may have, we re-evaluated existing concepts for analyzing the authentication process. The question we tried to answer was: {\em Are the methodologies described in the literature complete in the sense that there are not other options to attack \gls{oid}?}

\begin{figure}
        \centering
        \begin{subfigure}[b]{0.47\linewidth}
                \includegraphics[width=\textwidth]{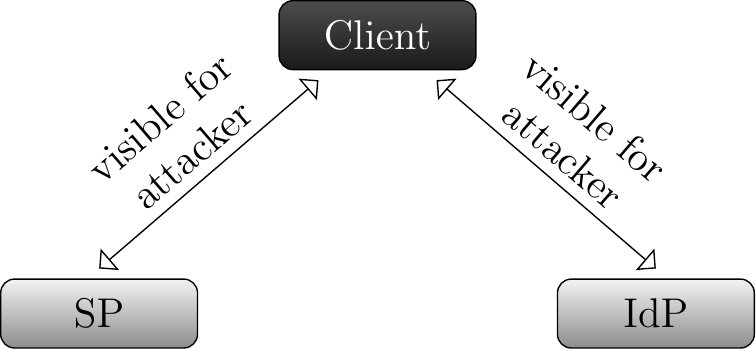}
                \caption{Old attacker paradigm.}
                \label{fig:model_old}
        \end{subfigure}%
        ~ %add desired spacing between images, e. g. ~, \quad, \qquad, \hfill etc.
        \begin{subfigure}[b]{0.47\linewidth}
                \includegraphics[width=\textwidth]{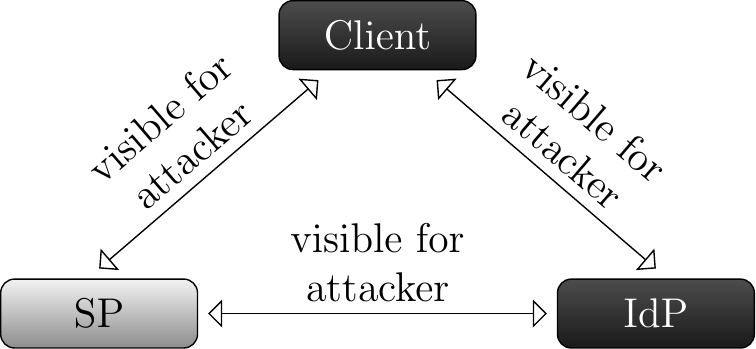}
                \caption{Our new attacker paradigm.}
                \label{fig:model_new}
        \end{subfigure}
        \caption{Our new attacker paradigm uses a \emph{malicious \gls{idp}} and can thus see all relevant messages.}
        \label{fig:model_comparison}
\end{figure}

\paragraph{New \gls{sso} Attacker Paradigm}
Since in \gls{oid} it is very easy for anyone to run an \gls{idp}, we extended the attack methodology and considered malicious \glspl{idp} as well. By running a malicious \gls{idp}, we enhance the attacker's capabilities: he is able to read and manipulate all messages exchanged between a honest \gls{sp} and the malicious \gls{idp}, even for messages that do not pass trough the browser (cf. \autoref{fig:model_comparison}). Thus, the attacker has better control over the \gls{sso} message flow, which results in a more thorough security analysis of \gls{sso}.

This novel approach for attacking an \gls{sp} revealed four new attack classes: \acrlong{trc}, \acrlong{ids}, \acrlong{kc} and \acrlong{ds}. 
All attacks work in the web attacker model or in even weaker variants. 
Thus the practical impact of these attacks is very high: For \acrlong{trc}, the \gls{oid} accounts of any user who visits the attackers web page can be compromised. 
For \acrlong{ids}, \acrlong{kc} and \acrlong{ds}, the effect is even more devastating: We can fully compromise \emph{all} accounts on an \gls{oid} \gls{sp}, \emph{without any user interaction}.

\paragraph{Methodology}
After an initial white-box analysis of the different \gls{oid} implementations, we used our own \gls{idp} to perform realistic black-box tests against running target \gls{sp} implementations.
For that purpose we used our automatic security testing tool \textbf{\gls{oida}} (Section \ref{sec:implementation}), allowing us to apply all presented attack classes on arbitrary \glspl{sp}.
The results of both analysis phases were then verified as follows:
We set up a victim account on each \gls{sp} implementation, and verified in each case that we could access this account through a second (attacker-controlled) browser, running on a different PC without the victim's credentials.

The validity of all attacks found has strictly been verified in the  \emph{Web attacker model}~\cite{webattackermodel}:
The attacker only controls the incoming and outgoing messages to and from web applications which he controls (e.g. malicious clients, \glspl{sp} and \glspl{idp}); all other network traffic is unknown to him.
He can also freely access victim web applications through their interface exposed in the WWW, through a web browser or a modified HTTP client.
An attack is considered successful if the attacker gets illegitimate access to protected resource at the legitimate \gls{sp}.

We do \emph{not} assume full control over the network, for instance, we do not use the (stronger) standard cryptographic attacker model, which yields weaker results.
Additionally, we do not consider phishing attacks -- the attacker does not imitate a legitimate \gls{sp} and we do not trick out a victim to use the attacker controlled \gls{idp}.

\paragraph{Results}
\label{sec:results}
We were able to find four novel attacks on \gls{oid}: \\

\begin{itemize}
    \item \textit{\acrlong{trc}} introduces an attacker acting as a malicious \gls{sp}.
         The attacker then forwards the received tokens to other \glspl{sp}.
     \item \textit{\acrlong{kc}} exploits a vulnerability in the key management implementation of the \gls{sp}, resulting in the use of an untrusted key. 
         The attacker acts as a malicious \gls{idp}.
    \item \textit{\acrlong{ids}} introduces an attacker in the role of a malicious \gls{idp}, generating tokens in the name of other (trusted) \glspl{idp}.
    \item \textit{\acrlong{ds}} exploits the usage of untrusted identities transmitted during the discovery phase by using a malicious \gls{idp}.
\end{itemize}

We evaluated these attacks against \evalnumber{16} implementations mainly taken from the official \gls{oid} Wiki~\cite{openidlibs}.
\autoref{tab:results} summarizes the results: were able to compromise \evalnumber{11} of them.
Our results show that the verification of a security token is a nontrivial task in \gls{oid}:
Dependencies between different data structures must be taken into account (e.g.\ \gls{asso} name and association key) and REST parameters must be checked with great care (Section \ref{sec:lessons_learned}).

%As a general result, it seems that it is better for the Service Provider to really "forget" all information received during steps (1.) to (3.) and start verification with data from Step (4.) only.

\paragraph{Responsible Disclosure}
All vulnerable projects have been informed and most acknowledged our findings.
In case we did not receive any reaction, we filed a CVE.
We cooperated by proposing and providing bug fixes, which were applied in some cases ~\cite{CVE-2014-1475,CVE-2014-2048,joidack,owncloudack,slashdotAck,CVE-2014-8249,CVE-2014-8250,CVE-2014-8251,CVE-2014-8252,CVE-2014-8253,CVE-2014-8254,CVE-2014-8265}.

\paragraph{Contribution}
\label{sec:contribution}
The contribution of this paper can be summarized as follows: \\
\begin{itemize}
    \item We propose a novel attacker paradigm for the analysis of \gls{sso} protocols -- the use of a malicious \gls{idp}. As a result, the security evaluation is more comprehensive. 
    \item We describe four novel attack classes on \gls{oid} by using a malicious \gls{idp}, all strictly in the Web attacker model.  These attacks provide novel insights into the problems of token verification for \glspl{sp}, and of enforcing the message flow intended by the \gls{oid} specification. 
    \item We give a systematic overview on \gls{oid} security and show that roughly 68\% of the analyzed implementations are vulnerable, including Sourceforge, Drupal and \gls{oc}. 
     \item We contribute to a better understanding of \gls{sso}, especially the trust establishment between \gls{idp} and \gls{sp} during the discovery and association phases in \gls{oid}.
    \item We develop \gls{oida}, a free and \gls{opensource} malicious \gls{oid} \gls{idp} capable of executing our novel and previous discovered attacks~\cite{openidattacker}.
%        Please note that although our attacker paradigm allows the attacker to setup a malicious \gls{sp} or act as a malicious \gls{sso} client for exploiting the vulnerabilities, it is only necessary to act as a malicious \gls{idp} to detect, whether the analyzed \gls{sso} implementation is vulnerable or not.
\end{itemize}

\paragraph{Outline}
% In the following section, we explain the benefit of using a malicious \gls{idp} for analyzing \gls{sso} protocols \todo{Evtl. raus falls die Section rausfliegt}.
% \autoref{sec:attack_model} presents the attacker model.
In the following section, we present the computational and security model and describe our new \gls{sso} attacker paradigm.
\autoref{sec:openid} introduces \gls{oid} and the protocol flow.
In \autoref{sec:analyzingmodel}, we elucidate the verification processes of authentication tokens and provide a general analysis for \gls{sso}.
In \autoref{sec:attackclasses}, we present novel attacks regarding \gls{sso}.
The methodology of analyzing \gls{sso} systems will be expounded in \autoref{sec:methodology}.
The results of the provided evaluation are supplied in \autoref{sec:evaluation}.
\autoref{sec:implementation} delineates the implementation of our tool, \gls{oida}.
Related work is discussed in \autoref{sec:relatedwork}. 
In \autoref{sec:lessons_learned} we sum up the lessons that we can learn from the paper. 
Finally, we conclude in \autoref{sec:conclusion}.

\section{Computational and Security Model}
\label{sec:attack_model}

\begin{figure}[!ht]
    \centering
    \includegraphics[width=\linewidth]{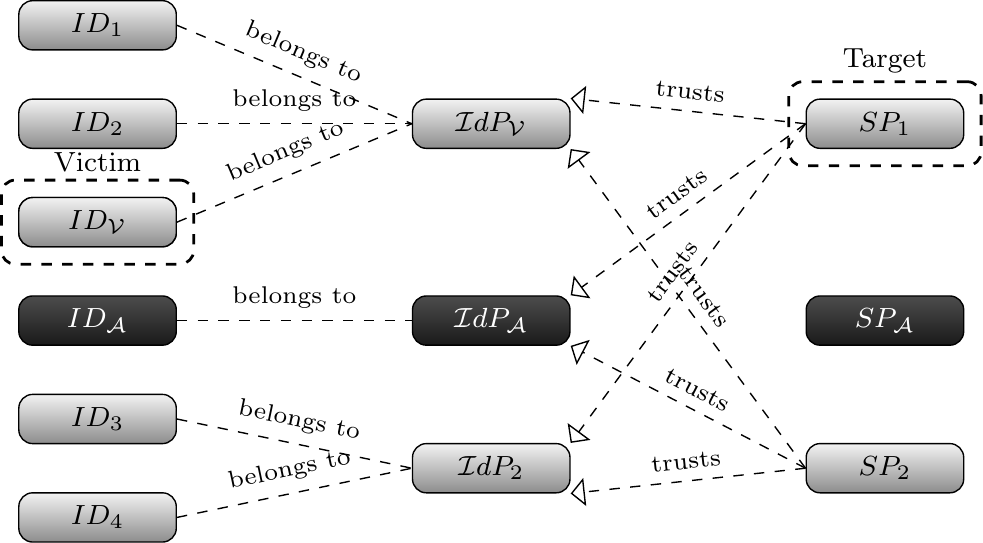}
    \caption{%
        \gls{sso} in the real world involves multiple clients, multiple \glspl{idp} and multiple \glspl{sp}.
        $SP_1$ can even \emph{trust} tokens of $\IdP_\A$, but only for its corresponding clients, i.e.\ $\ID_\A$.
    }
    \label{fig:attack_scenario}
\end{figure}

\paragraph{Computational Model}
\autoref{fig:sso_overview} shows a basic \gls{sso} login procedure. However, the real world is more complex and includes multiple clients, \glspl{sp}, and \glspl{idp}. \autoref{fig:attack_scenario} illustrates this scenario. Please note that while each \gls{sp} may trust several \glspl{idp}, and each \gls{idp} may serve many \glspl{sp}, each client's ID belongs to exactly one \gls{idp}.

In \gls{oid}, there is (in contrast to other SSO systems) an ``open'' trust relationship between  \gls{sp} and  \gls{idp}: The \gls{sp} trusts tokens created by any $\IdP$, as long as $\URLIdP$ is contained in the document retrieved from $\URLID_\C$. 
Thus it is easy to inject a malicious \gls{idp} $\IdP_\A$ into this ecosystem by simply creating a new (malicious) client ID where the discovery document points to $\URLIdP_\A$. 
Additionally, we can also run a malicious \gls{sp} $SP_\A$. 
Since we now control each type of communicating entities in an \gls{oid} system, we also control (and are thus able to modify) all types of messages. 
This is especially important in the \emph{Analyzing Mode} (cf. \autoref{sec:implementation}), where we modify certain parameters in each message type and test it against a honest instance of an \gls{sp}.

Please note that control over {\em all types of messages} should not be confused with control over {\em all messages}: As \autoref{fig:attack_scenario} shows, we cannot access messages exchanged between honest parties (e.g. $\IdP_2$ and $SP_2$).

\paragraph{SSO Attacker Paradigm}
The goal of the attacker is to access a protected resource to which he has no entitlement.
To achieve this goal, he may use the resources of a web attacker only: he can set up his own web applications and he can lure victims to them.
Furthermore, in three of four attacks described in this paper (\acrshort{ids}, \acrshort{kc} and \acrshort{ds}) the attacker is even more powerful: by using the malicious \gls{idp} only, the attacker can break into every \gls{oid} account on the target \gls{sp} without any victim's interaction. 
Thus, there is no possibility for the victim to detect or mitigate the attacks. %In summary, the attacker needs only one domain accessible on the Internet. 

% 
% The goal of the attacker is to access a protected resource to which he has no entitlement.
% To achieve this goal, he may use the resources of a web attacker only: he can set up his own web applications and he can lure victims to them.

% \begin{figure}[h]
%  \centering
%  \includegraphics[width=0.5\textwidth]{./img/AttackerModels.pdf}
%  % AttackerModels.pdf: 2818x1660 pixel, 72dpi, 99.41x58.56 cm, bb=0 0 2818 1660
%  \caption{The attacker can play different roles: Legitimate client, Malicious \gls{idp} and Malicious \gls{sp}. A Man-in-the-middle attacker eavesdropping packages between the victim and the \gls{sp} are not considered further.}
%  \label{fig:attackerroles}
% \end{figure}

In an \gls{sso} environment, the web attacker can play different roles (\autoref{fig:attack_scenario}):
\begin{inparaenum}
    \item {\bf Malicious client.} He can start an \gls{sso} session like any other client.
        Note that the attacker's identity $ID_\A$ belongs to $\IdPA$, but the victim's identity $ID_\V$ belongs to $\IdPV$.
    \item {\bf Malicious \gls{idp}.}
%        By acting as a malicious \gls{idp} we do not run phishing attacks and we do not try to steal the victim's login credentials, because the victim is associated with another legitimate IdP.
        The malicious \gls{idp} ($\IdPA$) in our model is able to generate valid as well as malformed authentication tokens (attack tokens).
    \item {\bf Malicious \gls{sp}.} %($SP_\A$).
        In our experiments, we never used any special properties of $SP_\A$: it is sufficient that the attacker just controls a domain ($\URLA$).
\end{inparaenum}

\section{OpenID: Technical Background}
\label{sec:openid}

\begin{overview}
    \begin{itemize}
        \item \emph{Background} appears more often than \emph{Foundations}
            \begin{itemize}
%                 \item Maybe: \emph{Technical Background} ?
            \end{itemize}
        \item SSO?
        \item OpenID
            \begin{itemize}
                \item Question: Relying Party or Service Provider (can be changed in glossaries.tex)
            \end{itemize}
    \end{itemize}
\end{overview}

\Gls{oid}~\cite{openid20} is one of the main \gls{sso} standards.
% \gls{oid} is decentralized and permits the usage of freely chosen \glspl{idp}.
In contrast to \gls{oauth} and \gls{saml}, where a trust relationship between an \gls{sp} and an \gls{idp} needs to be established beforehand, \gls{oid} does not require any registration or configuration at the \glspl{sp}.
%In this manner, \gls{oid} offers scalability, flexibility and simpler login procedure, since the user can just start the login process using an \gls{idp} of his choice.
% Noteworthy is the fact that the user's identity, certified by the \gls{idp}, is bound to an URL of the domain controlled by the \gls{idp}.
%
\gls{oid} is available as a module for \glspl{cms} like \gls{wp} and Joomla, or it is even directly shipped with the application, for example, with Drupal and \gls{oc}. 
Libraries for all commonly used Web programming languages are available~\cite{openidlibs}.
Millions of users already own an \gls{oid} as ,for instance, Google, Yahoo, and AOL automatically assign one to each user. % and according to~\cite{openidstatistics}, there are more than 12,000 websites that enabled \gls{oid} login.

\begin{figure*}[t]
    \centering
    \includegraphics[width=0.9\linewidth]{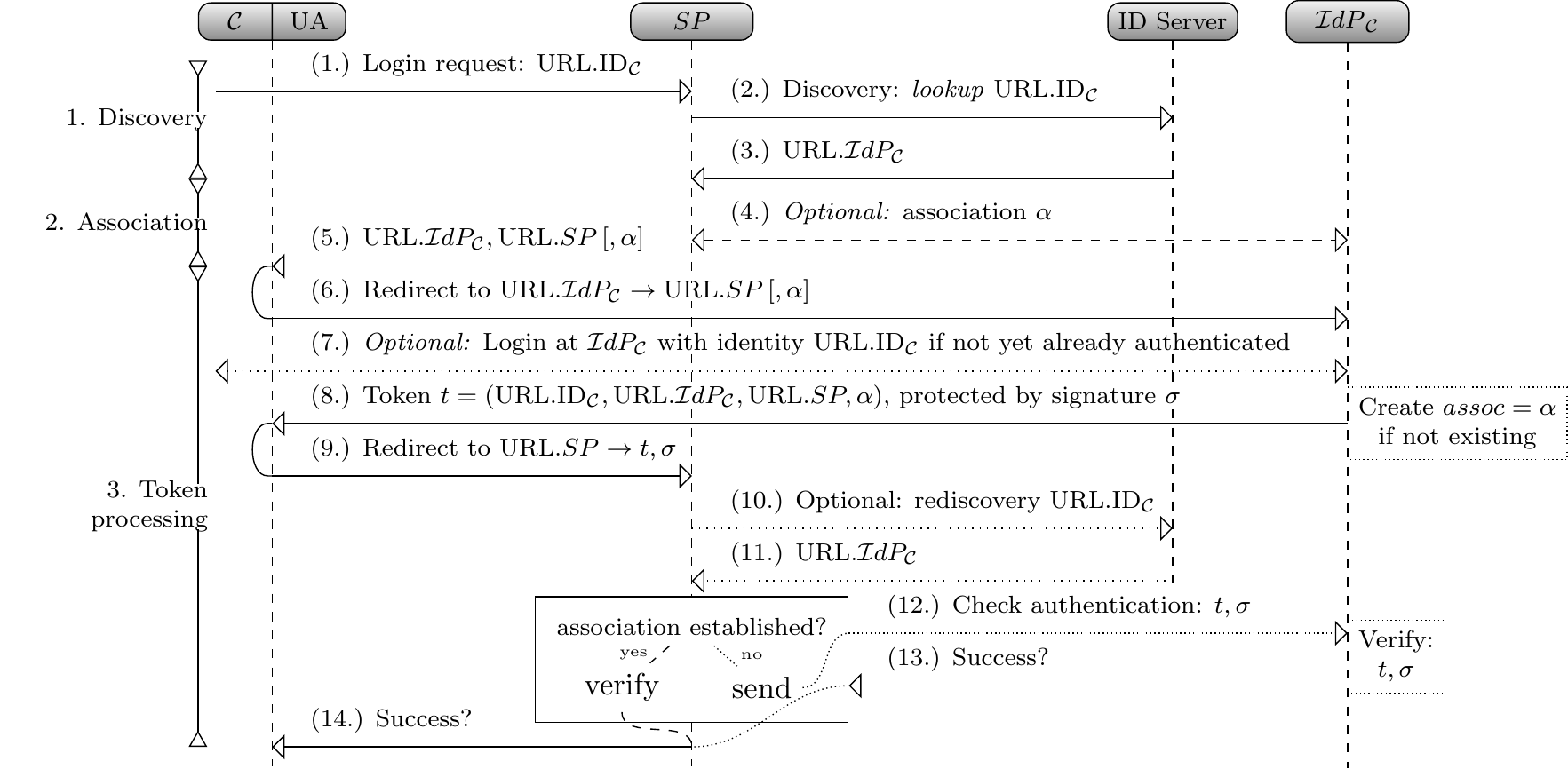}
    \caption{The \glsentrytext{oid} protocol flow.}
    \label{fig:openid_simple}
\end{figure*}

\paragraph{Notation} 
In \gls{oid}, an identity of a client $\C$ is represented by a URL\@.
Therefore, we define it as $\URLID_\C$.
Correspondingly, we define the URL of a client's \gls{idp} by $\URLIdP_\C$ and for an \gls{sp}, we use $\URLSP$.\footnote{$\URLID_\C$ and $\URLIdP_\C$ need not necessarily belong to the same domain.}

%\Gls{oid} uses a Hash MAC to protect the token, but the \gls{oid} specification~\cite{openid20} always claimed to it as a signature, so this paper will do this, too.

\paragraph{Protocol}
\Gls{oid} consists of three phases as shown in \autoref{fig:openid_simple}.
%\begin{figure}[!ht]
%    \centering
%    \includegraphics[width=0.9\linewidth]{tikz/openid_phases/pic}
%    \caption{The three phases of the \glsentrytext{oid} protocol.}
%    \label{fig:openid_phases}
%\end{figure}
In the \textbf{discovery} phase, the \gls{sp} collects information about $\C$'s requested identity ($\URLID_\C$) and determines $\URLIdP_\C$.
In the \textbf{\gls{asso}} phase, the \gls{sp} and the \gls{idp} establish a shared secret $\alpha$ intended to be used for signing and verifying the token.
The \textbf{token processing} phase then includes the creation of the token by the \gls{idp}, its transport to the \gls{sp} via $\C$'s \gls{ua}, and its verification by SP.
\autoref{fig:openid_simple} describes the \gls{oid} login process more precisely: \\

\begin{enumerate}
    \item $\C$ wishes to access a resource at the \gls{sp} and enters his identity $\URLID_\C$.

    \item The \gls{sp} then starts the discovery by requesting the document at $\URLID_\C$.
        %The ID server may use the same domain as the \gls{idp}, or it can be a totally different entity.

    \item A document containing $\URLIdP_\C$ is returned.
        % response can be either an HTML or an \gls{xrds} document~\cite{xrisyntax20}.
        % \todo{\gls{xrds} erklären? Normalerweise wird im Falle von HTML nochmal ein \gls{xrds} Dokument angefragt. Erwähnen?}
        % For simplicity, this is in both cases an XML document that contains mainly the endpoint URL of $\C$'s \gls{idp}, for example, \texttt{https://www.google.com/accounts/o8/ud}.
        % Optionally, it can also include $\C$'s \gls{oid} URL at his \gls{idp}.
        % This is a feature of \gls{oid} and allows a client to use its own \gls{oid} identifer, for example, \texttt{http://mysite.com/me}, which then points to a Google identity, for example, \texttt{http://google.com/me}.
       % $\C$ can then easily change its \gls{idp} at one place, i.e.\ at \texttt{http://mysite.com/me}.
        % The \gls{sp} extracts the included information, for example, the \gls{idp} endpoint URL\@.

    \item Using $\URLIdP_\C$, the \gls{sp} can establish an \gls{asso} with the \gls{idp}.
        This %optional step 
is basically a \glsfirst{dhke} to establish a \emph{shared secret} $s$.
        Additionally, the \gls{idp} freely chooses a string $\alpha$ that is used as a name for the association. 
        It is used to reference the key material $k$ derived from $s$ on both sides, and has an expiration time.
        Note that in this phase, the \gls{sp} and the \gls{idp} are directly communicating with each other, which means that a  web attacker cannot interfere with this communication.

    \item Afterwards, the \gls{sp} has all necessary information to validate an \gls{oid} token created by $\IdP_\C$. % in the course of all later steps. % to later validate  an \gls{oid} token.
        It responds to $\C$'s initial login request of Step (1.) and sends an \emph{authentication request} containing $\URLIdP_\C$, $\URLSP$ and optionally $\alpha$.%, in case that an \gls{asso} has been established.

    \item $\C$ is redirected to $\URLIdP_\C$.

    \item If $\C$ is not yet logged in, he must authenticate to $\IdP_\C$.

    \item $\IdP_\C$ creates a token $t$ for $\C$ containing $\C$'s identity $\URLID_\C$, its own URL address $\URLIdP_\C$ and  $\URLSP$. 
%It is saved under a reference value $\alpha$ and also contained in $t$.
        $\IdP_\C$ then generates a signature $\sigma$ for $t$ using the key referenced by $\alpha$.
        Message (8) is called the \emph{authentication response} and is sent as an HTTP redirect to $\URLSP$.

    \item The authentication response is forwarded to the \gls{sp}.
    \item[(10.)-(11.)] The \gls{sp} can optionally start a rediscovery, for example, if it has not cached the previous discovery, cf.\ Step (2.)-(3.).
%$\C$ is redirected with $t$ and $\sigma$ to the \gls{sp} using $\URLSP$ contained in $t$.
%        Then, there are two possibilities:
%        If the \gls{sp} has established an \gls{asso} in Step (4.), it is able to verify the token without interacting with the \gls{idp}.

%    \item Otherwise, the \gls{sp} forwards $t$ and $\sigma$ to $\IdP_\C$ by using $\URLIdP_\C$ contained in the token.
%
%    \item $\IdP_\C$ verifies $\sigma$ using the private key identified by $\alpha$, which is contained in $t$, and responds whether the signature is valid or not.

    \addtocounter{enumi}{4}
    \item If the signature is valid, the \gls{sp} will map $\URLID_\C$ to a local  identity and respond accordingly to $\C$.
\end{enumerate}

\paragraph{Direct Verification} 
Establishing an association is optional according to the \gls{oid} standard. If the communication (4.) is missing, the authentication request does not contain $\alpha$, and no \emph{shared secret} was established with $\IdP_\C$. 
In this case, the \gls{idp} generates a fresh key and signs the token with it.
In this case, the \gls{sp} will not be able to verify the authenticity of the token by itself. 
Instead, it must send the token directly to the \gls{idp} in Step (12.), and accepts the result of the verification from Step (13.). 

\paragraph{Discovery in Detail}
% \begin{overview}
    % \begin{itemize}
        % \item Step (3.) reveals $\URLIdP_\C$.
            % \begin{itemize}
                % \item XRDS or HTML
            % \end{itemize}
        % \item Note that Step (5.) does not contain $\URLID_\C$. 
        % \item It is not necessary because $\C$ has to authenticate to $\IdP_\C$ so that $\IdP_\C$ know $\URLID_\C$.
        % \item Additionally, it can optionally contain a second identity $\URLIdP_{\C'}$.
    % \end{itemize}
% \end{overview}
To receive $\URLIdP_\C$ in Step (2.), the \gls{sp} fetches the document at $\URLID_\C$ (e.g. \url{http://myserver.org}).
This can be either an HTML or an \gls{xrds} document.
\autoref{lst:html_discovery} shows a minimal HTML document.

\begin{lstlisting}[language=html,caption={Minimal HTML discovery document.},label={lst:html_discovery}]
<html><head><title/>
<link rel="openid2.provider" 
     href="https://myidp.com/" />
</head><body/></html>
\end{lstlisting}

The element \lstinline[language=html]{<link/>} contains $\URLIdP_\C$ within the \texttt{href} attribute.
\Gls{xrds} documents contain the same information, but stored in XML data format.

Note that Step (5.) of the protocol does not contain $\URLID_\C$. 
This is not necessary, since $\C$ must authenticate to $\IdP_\C$. Consequently, $\IdP_\C$ knows the value of $\URLID_\C$.
However, the discovered document in Step (3.) allows optionally to include a second ``local'' identity ${\URLID_\C}^*$ (the value of the href attribute in \autoref{lst:html_discovery_id}):

\begin{lstlisting}[language=html,caption={$\C$'s identity stored in an HTML document.},label={lst:html_discovery_id}]
<link rel="openid2.local_id" 
     href="https://myidp.com/bob" />
\end{lstlisting}

If this is the case, steps (5.) and (6.) will include this value as well and $\IdP_\C$ is asked to use ${\URLID_\C}^*$.
This is, for example, useful if $\C$ owns multiple IDs at $\IdP_\C$.

\section{SSO Token verification}
\label{sec:analyzingmodel}

Token verification at the \gls{sp} is the most critical part within the \gls{sso} process.
It consists of many steps in order to guarantee the validity of the authentication. %, see \autoref{fig:analyzing_model_overview}.
%and verification of the cryptographic signature is only one of these steps.
%
%%During the research we investigated these steps irrespective of the used protocol and categorized them as shown in .
%
%% \begin{figure}[!ht]
%    % \centering
%    % \includegraphics[width=0.9\linewidth]{tikz/analyzing_model_overview/pic}
%    % \caption{Steps to process an \glsentrytext{sso} token.}
%    % \label{fig:analyzing_model_overview}
%% \end{figure}
%
This observation holds for \gls{sso} in general (\gls{saml}, \gls{oauth} and \gls{oid}).
In the following, these verification steps are discussed.

\paragraph{Message Parsing}
Each token has a specific structure. %, regardless whether this is \ XML, JSON\@ or a set of HTTP headers.
%\todo{Message parsing and Schema Validation? Mehr auf parsing eingehen}
%To provide conformity and interoperability, \gls{sso} protocols follow a clearly defined schema for every protocol message.
%It starts with the token's message format, for example, XML or JSON\@.
%Furthermore, the schema describes the token's layout, semantics, and the content of a message so that it can be evaluated during the verification process.
For instance, each \gls{oid} parameter starts with \texttt{openid.*}, and the required set of parameters must be checked by each application.
At the beginning, whenever an \gls{sp} receives a message, it has to be parsed into a data object so that it can be processed further. 
Any error during this parsing directly affects \gls{sso} security: for instance, if some data element is present twice with different content, the second content may overwrite the first during the parsing, or vice versa.
%Depending on the message format, different rules and operations for the conversion of the message are applied.
%Afterwards, the validity of all parameters, their values and format must be verified during this transformation.
Additionally, all required parameters must be present.

% Based on the results of past researches and their impact regarding the security\todo{@Vladi: Quelle?}, the execution of the parsing procedure and the schema verification is a critical step within the authentication process.
% Therefore, it should be implemented by taking into account all security related aspects.\todo{Christian: Irgendwie finde ich diesen ganzen Absatz unnötig}

\paragraph{Freshness}
Freshness of authentication tokens is important for preventing replay attacks.
It can be realized with two parameters:
\begin{inparaenum}
    \item a nonce, which is a random value selected by the \gls{sp} and/or
    \item a timestamp which defines the token's creation time or period of validity, and which is usually selected by the \gls{idp}.
\end{inparaenum}
%contained in every token\todo{(\emph{in every token} streichen? Es könnten ja auch nur timestamps sein, oder?)}.
%Its uniqueness allows an unambiguous token identification so that its reuse can be detected and prevented.
%Depending on the \gls{sso} protocol, its implementation and the concrete use-case,
%the lifetime can be explicitly defined by setting a token expiration time, or it can be implicitly configured by only using the token creation time.
%Lifetimes can differ from a few seconds up to some minutes or even several hours.
%With respect to security, the reuse of authentication tokens should be prevented, in order to mitigate replay attacks.
%% Therefore, all relevant attributes have to be verified.
%Note that not all \gls{sso} protocols deny the reuse of a token, for example, in \gls{saml} the token may explicitly contain the \texttt{<OneTimeUse/>} condition to enforce it.
\gls{oid} uses the parameter \texttt{openid.response\_nonce}.
It contains the creation time of the token concatenated with a random string.

\paragraph{Token Recipient Verification}
A token $t$ is intended for a single  \gls{sp}.
Thus, it should be guaranteed that (1.) $t$ can be successfully verified by a single \gls{sp} only, and (2.) that $t$ is delivered to the correct \gls{sp}.
%
%(1.) Every \gls{idp} supports multiple \glspl{sp} and creates authentication tokens for each of them.
%From a security point of view, it is crucial to bind the use of a specific token to the according \gls{sp}, otherwise the attacker acting as a malicious SP could reuse all received tokens to gain access at the target SP.
OpenID uses the $\URLSP$ parameter for purpose (1.). This parameter should be checked by the SP.
%because every \gls{sp} is an interdependent \todo{(Independent?)} instance.
% If a client asks his \gls{idp} to create an authentication token for $SP_1$, it must not be possible to use it on $SP_2$.
%\Gls{sso} protocols provide such bindings by means of different parameters, for example, the \texttt{Recipient}-attribute in \gls{saml},  in \gls{oid} and the \texttt{redirect\_uri}-parameter in \gls{oauth} 2.0.
%In most cases, the type of these attributes is a URI, because URIs can be used as a unique identifier for the concrete \gls{sp}.
%In this manner, tokens issued for one \gls{sp} should not be redeemable on another one.
%
%An applicable attack scenario is the usage of a compromised or malicious \gls{sp} by an attacker in order to gain access to authentication tokens.
%In the following step, the attacker forwards the obtained tokens to another \gls{sp}, which is the victim of this attack.
%If the validation of the recipient attributes is not correctly performed or, even worse, there is no validation at all, the tokens will be accepted and the attacker impersonates the victim.
% An erroneous verification may consist, for example, in checking if the recipient attribute partially matches to the expected value.
%In this manner, the validation can be bypassed by using subdomains or specially chosen URIs. \todo{Christian: Ist das hier nicht schon ein zu konkreter Angriff? Oder passt das so?}
%
For (2.), the HTTP-Receiver of the redirect message sent by the \gls{idp} is given in the \gls{oid} parameter $\URLSP$. 
Here the \gls{idp} must check that this parameter is valid. % for the targeted SP.

\paragraph{\glsentrytext{idp} Verification}
%In \gls{sso} the \gls{idp} is responsible for the user authentication and the generation of authentication tokens, which will be sent to the \gls{sp}.
The \gls{sp} receiving a token should verify: (1.) the origin of the received token and (2.) the validity of the statements contained.
(1.) is verified in three steps:
(1.1) The \gls{sp} must determine the unique identity of the \gls{idp} (e.g.\ an URL) which issued the authentication token.
(1.2) The \gls{sp} must fetch the corresponding key material associated to that identity.
(1.3) Using this key material, the signature of the token is verified.
%%In most cases, the \gls{sp} stores the keys locally for all known \glspl{idp} and thus, the \gls{sp} has just to determine the correct one.
%Afterwards, in (1.2), the \gls{sp} proves if the key used for protecting the authentication token matches to the determined key in (1.1).
%In this manner, the \gls{sp} can ensure that the received token was generated and signed by a trustworthy \gls{idp} and the token can be further processed.
%In the case of \gls{saml}-based \gls{sso}, the trust relationship is provided via the exchange of metadata~\cite{SAMLMeta} containing the key material.
%For \gls{oid}, \gls{sp} and \gls{idp} establish a shared secret via \gls{dhke} used for protecting and verifying the token.
%In summary, the relation between the \gls{idp}'s identity and the stored key material is crucial for the security and must be implemented correctly.
%
In (2.) the \gls{sp} should verify whether the \gls{idp} is allowed to make the statements in the token, for example, $\IdPA$ must not issue tokens in the context of $\IdPV$.
%In \gls{oid}, a user's identity is represented by a unique URL controlled by exactly one \gls{idp}.
%Therefore, a second entity should not be able to issue tokens containing the same identities URI, which will be successfully verified.
%
%With respect to security, an inaccurate implementation of this verification step allows an adversary to issue tokens in the name of other \glspl{idp} and sign them with key material controlled by himself (and not by the correct \gls{idp}).
%If this is the case, the tokens would be successfully verified and processed by the \gls{sp}.
%Consequently, the impersonation of any user on the affected \gls{sp} is possible.

%\paragraph{Verification of additional conditions} The token may contain additional condition which must be verified.

\paragraph{Cryptographic Token Verification}
%Finally, the integrity of the received authentication token must be verified.
%In most cases, this is a signature or a MAC, but for a better readability, we use the terms \emph{sign} and \emph{signature}.
%The verification process 
For step (1.3) above, the signed parts must be determined. 
The \gls{sp} must be able to distinguish signed from unsigned parts within the token.
For instance, in \gls{oid}, it should be able to distinguish signed HTTP header fields from unsigned ones.
Additionally, it should check if all parameters that are required to be signed are indeed signed.\footnote{In the context of \gls{saml}, this has been shown to be quite challenging~\cite{somorovskySAML}.}
%Such parameters contain sensitive information and thus must be protected.

For step (1.2) above, the right keys must be chosen.
The \gls{sp} uses the key material associated with the selected \gls{idp}.
%This step is closely related to \emph{\gls{idp} verification}.
If this \gls{asso} between key material and identity can be overwritten (cf. \autoref{sec:keyconfusion}), novel attacks are feasible.
%In  we introduce a novel attack technique allowing to bypass the verification logic by loading the wrong key.
%As a result of the attack, authentication tokens signed with an unrelated key will be successfully verified.
%Consequently, unauthorized access is feasible.

%In (3.) the verification of the signed parts will be carried out according to the selected cryptographic mechanism.
%In case of \gls{saml}, this is a digital signature, and for \gls{oid}, HMACs are applied. (3.) relies on the security of the chosen cryptographic algorithm.
%Unfortunately, attacks exploiting known weaknesses of these algorithms or their implementation can lead to serious security vulnerabilities~\cite{Tsudik92messageauthentication,duong2009flickr,XMLDigitalSignatures2013}.
%
%Finally, in (4.) the content of the token is forwarded to the application logic.
%It is crucial to verify that the verification and application logic process the same data.
%Based on past researches, such attacks make unauthorized access applicable~\cite{somorovskySAML}.

% With respect to this step Somorovsky et.al. provided a security evaluation of \gls{saml}-based interfaces revealing novel techniques to bypass the integrity protection of authentication tokens in 2012~\cite{somorovskySAML}.
% The authors exploited the separation of the two logics, verification and application logic, in order to inject malicious content.

\section{Novel Attacks}
\label{sec:attackclasses}

In this section, we give generic descriptions of four novel attacks on \gls{oid}, which are effecive against different  implementations of OpenID (cf. \autoref{tab:results}).
The first two attacks, \acrlong{trc} and \acrlong{kc}, are protocol independent and can be applied to other \gls{sso} protocols. \acrlong{ids} and \acrlong{ds} exploit characteristic of \gls{oid}.
% Three of them are novel.
% They make use of a malicious \gls{idp} which is not assumed in any previous work.
% For \gls{trc}, we adopted the idea of a \gls{saml} attack.

%-----------------------------
\subsection{\acrlong{trc}}
\label{sec:recipientconfusion}
\begin{figure}[!ht]
    \centering
    \includegraphics[width=\linewidth]{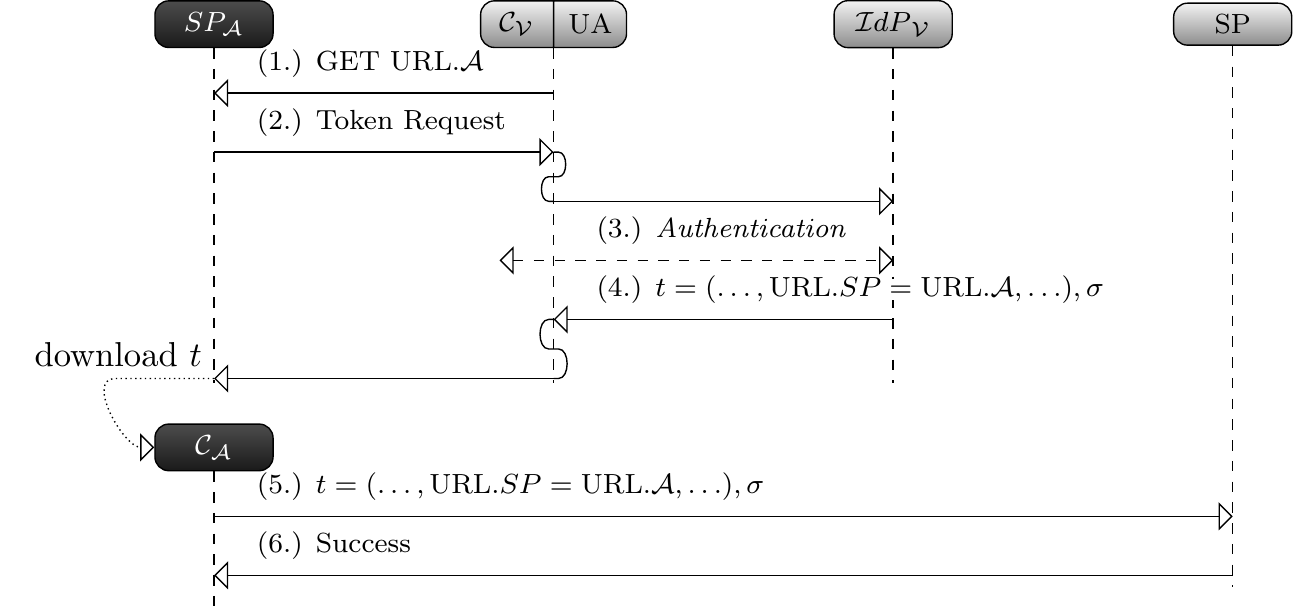}
    \caption{\acrlong{trc} Attack.}
    \label{fig:attack_returnto}
\end{figure}

\emph{\gls{trc}}  attacks as shown in \autoref{fig:attack_returnto} target a missing $\URLSP$ parameter verification. This violates
condition (2.) of the \emph{token recipient verification} step (cf. \autoref{sec:analyzingmodel}).

%\Gls{trc} contains two phases -- detection and exploit. For the detection we use our \gls{idp} ($\IdPA$), whereas for the exploit we need a website visited by $\C_\V$\footnote{This condition perfectly matches the web attacker model.}.

\emph{Detection phase.} The attacker uses $\IdPA$ and generates tokens containing identity $ID_\A$. Additionally he sets the value of $\URLSP$ to an arbitrary URL (different from the URL of the target SP) and sends the token to the target \gls{sp}. Finally, he observes the behavior of the target \gls{sp}: If the \gls{sp} accepts the token, then the value of $\URLSP$ is \emph{not} validated, and \gls{trc} is applicable.

\emph{Exploit phase.}
In order to exploit the vulnerability, the attacker $\A$ sets up a web application running on $\URLA$ (e.g. a weather forecast service), to initiate an OpenID authentication and to collect authentication tokens.
The exact protocol flow is shown in \autoref{fig:attack_returnto}.

\begin{enumerate}
\item The victim client ($\C_\V$) accesses the web application deployed on $\URLA$.
\item The attacker creates a \emph{Token Request} containing $\URLSP = \URLA$.%
\item $\C_\V$ authenticates to $\IdPV$. If he is already authenticated, this step is skipped.
\item $\IdPV$ generates the token $t$ and sends it back to $\C_\V$, with a redirect to $\URLSP = \URLA$. The client's \gls{ua} executes this redirect, and thus sends the token to $\A$.
\item Finally, $\C_\A$ downloads the collected token $t,\sigma$ from $SP_\A$ and uses it to log in on the target \gls{sp}.
\end{enumerate}

Note that in case that $\C_\V$ is already authenticated to $\IdPV$, Steps 2,3,4 and 5 will be executed without any user interaction.

\Gls{trc}  is a generic attack and can be adapted to other \gls{sso} protocols like SAML and OAuth, since these include parameters similar to $\URLSP$. 
For \gls{saml}, this is the \emph{AssertionConsumerServiceURL} parameter~\cite[Section 3.4.1]{SAMLCore} which is already evaluated in~\cite{saas_ccsw14}.
In OAuth, the parameter is called \emph{redirect\_uri}~\cite[Section 4.2.1]{rfc6749}.

To mitigate the \gls{trc} attack, the \gls{sp} should verify whether the $\URLSP$ parameter contained in $t$ matches its own URL\@.
%If this verification is missing, the token is accepted and the attacker will be authenticated as the victim.

%------------------------------------
\subsection{\acrlong{kc}}
\label{sec:keyconfusion}

\emph{\gls{kc}} is a generic and very complex attack on \gls{sso}.
A detailed example is shown in \autoref{fig:attack_drupal}.
The goal of the attacker is to force the target \gls{sp} to use a key of the attacker's choice to verify a (forged) token $t^*$.
To achieve this goal, he must agree a common secret key (or a public key) with the target \gls{sp}, and thus play the role of a (malicious) \gls{idp}.
Then he may follow one of two strategies to succeed.
\gls{kc} attacks address the second part of the {\em cryptographic token verification} step (cf. \autoref{sec:analyzingmodel}).

\paragraph{Strategy 1} \emph{Overwriting the secret key handle of a trusted \gls{idp}.}
In the case of \gls{oid}, the key material is referenced by the \gls{asso} handle parameter $\alpha$.
Since the value of $\alpha$ is chosen by the \gls{idp} (and not by the \gls{sp}),
the attacker (acting as a malicious \gls{idp}) is able to set $\alpha$ to the same value as defined by the valid \gls{idp} in order to overwrite it with its own key values.
The attacker may get to know the original $\alpha$ by starting an attempt to log in as the victim on the target \gls{sp}.
He will then receive $\alpha$ in message (5.) of \autoref{fig:openid_simple}.

\paragraph{Strategy 2} \emph{Submit attacker's own key handle for signature verification.}
The \gls{asso} $\alpha$ is also part of the signed token $t^*$.
Thus, some \gls{sp} implementations are tempted to use this value to verify the signature.
The fact that the token may be issued by a malicious \gls{idp} clearly shows that this leads to a critical vulnerability:
If a malicious $\IdPA$ issues the token $t^* = (\URLID_\V,\URLIdP_\V,\URLSP,$ $\beta)$ and protects it with a signature key related to $\beta$, the target \gls{sp} may accept this token. 
Although $\IdPA$ is not entitled to issue such tokens.
This behavior is not clearly prohibited:
According to the \gls{oid} specification~\cite[Section 11.2]{openid20}, an \gls{sp} should verify that the discovered information (user's identity and \gls{idp}'s URL) maps the presented content in the received token.
Unfortunately, this check does not verify that the key used for signing the token belongs to the discovered \gls{idp}.

The idea of \gls{kc} can be adapted to other \gls{sso} protocols using digital signatures for integrity protection, for example, SAML~\cite[Section 4.4.2]{SAMLSecurity}.

\subsection{\acrlong{ids}}
\label{sec:idspoofing}

\emph{\gls{ids}} is an \gls{oid} specific attack.
Its goal is to create a token $t^*$ containing the victim's identity ($\URLID_\V$) by using the attacker's \gls{idp}.
It is successful if the target \gls{sp} accepts $t^*$.
Given the simplicity of this attack it is surprising that it has not been described before.
\gls{ids} attacks target condition (2.) of the \emph{\gls{idp} verification} step (cf. \autoref{sec:analyzingmodel}).

In \gls{oid}, a user's identity is represented by  $\URLID_{\V}$, which is controlled by exactly one \gls{idp} ($\IdPV$ with $\URLIdP_\V$).
Consequently, an \gls{idp} can make statements only for user identities bound to its domain.
Thus, $\IdPA$ should in theory not be able to create a valid token $t^*$ containing $\URLID_{\V}$.
For \gls{oid}, the corresponding check should work as follows:
According to the specification~\cite[Section 11.2]{openid20}, an \gls{sp} should start a (second) discovery on the  identity $\URLID_{\V}$ contained in $t^*$.
In this manner, \gls{sp} can discover whether $\URLID_\V$ belongs to the \gls{idp} contained in $t^*$, i.e.\ $\IdPA$ in this case.
If this step is not implemented properly, an attacker is able to inject identities, which are not controlled by his malicious \gls{idp}.
In this manner, the attacker can impersonate users with different, trustworthy \glspl{idp}, for example, Google or Yahoo, by using only his own $\IdPA$.

\subsection{\acrlong{ds}}
\label{sec:discoveryspoofing}

\emph{\gls{ds}} is an attack which is only possible if the \gls{sp} uses the second ``local'' $ID_\C^2 = {\URLID_\C}^*$ for identifying the client.
The \gls{oid} specification allows this usage of identity $ID_\C^2$ returned by the ID server~\cite[Section 10.1]{openid20}.
%
%% Another related attack to \gls{ids} is \gls{ds}.
%The concept of this attack addresses the idea of \gls{oid} to separate the URL submitted in the login form ($\URLID_\C$), from the concrete URL at the client's \gls{idp}.
%
%The original idea behind this concept is that a client can use his self-controlled domain, $\URLID_\C$, as an ID Server during the discovery phase by providing an \gls{xrds}/HTML document, see \autoref{lst:html_discovery}.
%The \gls{sp} will fetch this document and extract the user's real identity $\URLID_{\C^{*}}$ contained in it.
%Then, the \gls{sp} will redirect the user to its \gls{idp} ($\URLIdP_\C$), for example, Google, which generates the token.
%
The attack exploits the fact that this second $ID_\C^2$ cannot be used for discovery: Only the first $ID_\C^1$  uniquely determines the trusted \gls{idp}.
The attack can be outlined as follows (a detailed description is given in \autoref{sec:owncloud}):

\begin{enumerate}
    \item The attacker stores a (malformed) XRDS/HTML document on his ID Server containing the victim's second identity $ID_\V^2 = \URLID_\V$ (see \autoref{lst:html_discovery_id}).
    This document can be retrieved through $ID_\A^1 = \URLID_\A$.
    \item The XRDS/HTML document thus retrieved points to an \acrlong{idp}  $\IdPA = \URLIdP_\A$ under the control of the attacker.
    \item $\IdPA$ issues a valid token $t$ for $ID_\A^1$, and the target SP successfully verifies $t$.
    To match this verification to a local identity, the attacker must either perform another (second, optional) discovery using $ID_\A^1$, or retrieve the result of the first discovery. In both cases, he will get the local ID $ID_\V^2$, and consequently grant access to $\A$.
\end{enumerate}
\hspace{1pt}

%\gls{ds} is a surprisingly detected attack and .

\begin{overview}
    \begin{enumerate}
        \item Schema Validation
            \begin{itemize}
                \item Required parameters
                \item Well-formed attributes, e.g malformed signature: twitter attack?
            \end{itemize}
        \item Nonce verification
            \begin{itemize}
                \item \textcolor{red}{Replay-Attacks?}
            \end{itemize}
        \item Return\_to verification
            \begin{itemize}
                \item \textcolor{red}{Recipient Confusion (ACS Spoofing)}
            \end{itemize}
        \item \gls{idp} Verification
            \begin{itemize}
                \item[$\Rightarrow$] Re-Discovery (local/remote)
                    \begin{itemize}
                        \item[$\Rightarrow$] YES (remote)
                            \begin{itemize}
                                \item \textcolor{red}{Discovery Spoofing}
                            \end{itemize}
                        \item[$\Rightarrow$] YES (local)
                            \begin{itemize}
                                \item \textcolor{red}{Key Confusion ($\$SESSION$ Spoofing)}
                            \end{itemize}
                        \item[$\Rightarrow$] NO
                            \begin{itemize}
                                \item \textcolor{red}{ID Spoofing}
                            \end{itemize}
                    \end{itemize}
                \item[$\Rightarrow$] Endpoint Verification
            \end{itemize}
        \item Signature Verification
            \begin{itemize}
                \item association = getAssociation(assoc\_handle)
                    \begin{itemize}
                        \item \textcolor{red}{Association Spoofing}
                    \end{itemize}
                \item association = getAssociation(\$service[url],assoc\_handle)
		\item Signature algorithms: MD5 length-extension attack
            \end{itemize}
    \end{enumerate}
\end{overview}

\section{Methodology}
\label{sec:methodology}

% The evaluation results are summarized in~\autoref{tab:results}.
% We categorized them in dependence of the according \gls{oid} implementations and validated each target against the defined attack classes.
% The last column summarizes whether unauthorized access is possible.

%-----------------------
\paragraph{Target \glspl{sp}}
We selected \evalnumber{15} \gls{opensource} implementations including libraries and frameworks that support \gls{oid}, mainly taken from the official \gls{oid} website~\cite{openidlibs}\footnote{Note that some of the libraries are listed multiple times, for example, libopkele is the module used in Apache mod\_auth\_openid, the listed Python Django \gls{oid} framework uses janrain etc.}.
%We did not focus on a specific programming language as our attacks abuse logical flaws in the protocol.
%Since information flow flaws are independent of programming languages, w
We tried to cover every available language:
Our list contains implementations in
.NET,
C++,
ColdFusion,
Java,
JavaScript,
Perl,
\gls{php},
Python,
and Ruby.
% In the case of janrain, there are three independent implementations in \gls{php}, Python and Ruby.
We added Drupal to the target list, since it is a widely used \gls{cms} and has a custom implementation of \gls{oid}.
%Other \glspl{cms} like \gls{wp} or Joomla rely on third party libraries -- those two examples use janrain, which is already contained within the list.
The only implementation that did not permit a white-box analysis is Sourceforge~\cite{sourceforge}.
We included it because it is a very prominent site supporting \gls{oid} (Alexa~\cite{alexa} rank 160) and because it does not use one of the inspected implementations listed on~\cite{openidlibs}.

%-----------------------
\paragraph{White-Box Tests}
We used white-box tests to analyze the source code and the protocol flow of each target.
%To this end, we used our own custom \gls{oid} \gls{idp} -- \gls{oida}.
%The tool is free, \gls{opensource} and can be downloaded on~\cite{openidattacker}.
Based on the white-box tests, we developed the concepts for the attack classes described in \autoref{sec:attackclasses} and implemented them in \gls{oida}.

%-----------------------
\paragraph{Setup}
For each implementation, we created a working virtual web server/virtual CMS server, and deployed the framework in it.
For Sourceforge, we used the live website.

\begin{figure}[!ht]
    \centering
    \includegraphics[width=\linewidth]{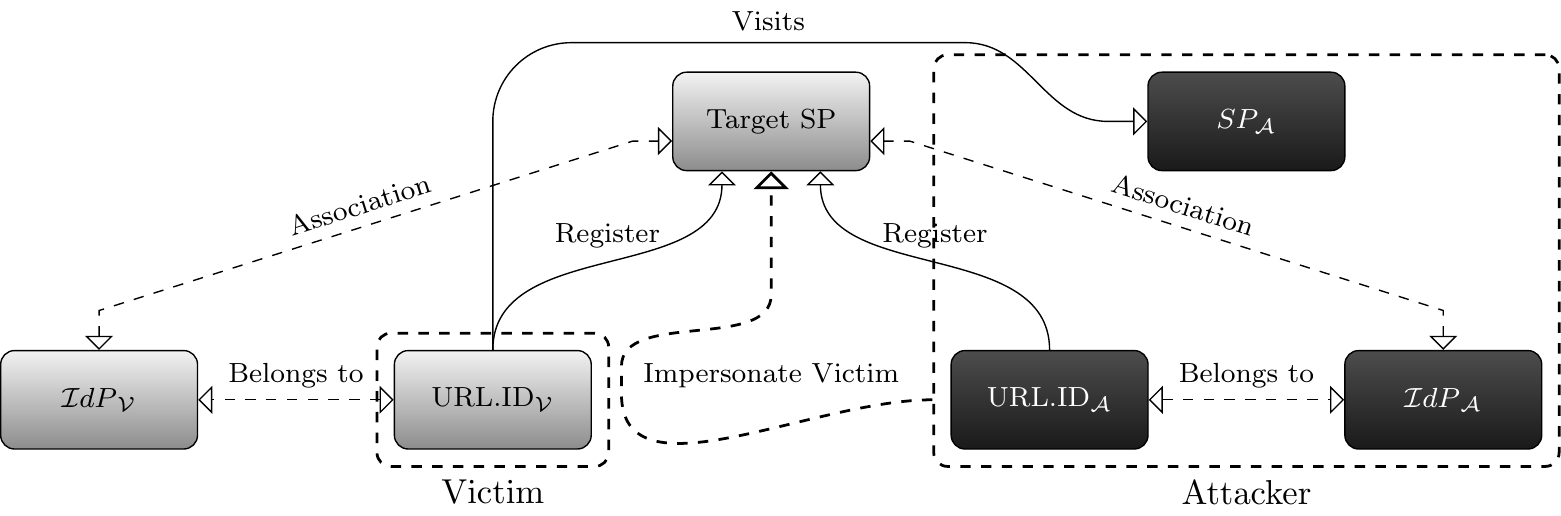}
    \caption{Evaluation setup and goal.}
    \label{fig:evaluation_setup}
\end{figure}

We registered two accounts on each target as shown in \autoref{fig:evaluation_setup}:
As victim $\V$, we used an account at a trusted \gls{idp} %like Google or Yahoo 
to register a local account on the target \gls{sp}.
Using a second browser on a different PC %(for isolation of both authentication processes) 
we registered a second account for $\A$ at the target \gls{sp}, associated with an account on our custom malicious \gls{idp} -- the \gls{oida} account.

In this step, the second account was mainly used to verify that the \gls{oida} \gls{idp} is working flawlessly and that the target is able to verify valid tokens created by our tool.

%-----------------------
\paragraph{Black-Box Test using \gls{oida}}
%\gls{oida} is our custom implementation of an \gls{oid} \gls{idp}.
Offering the ability to manipulate each parameter in every phase of the \gls{oid} protocol, \gls{oida} (cf. \autoref{sec:implementation}) allows to evaluate the token verification of an \gls{sp} in a very flexible way.
The tokens to attack the target \gls{sp} are created automatically. % and sent to it by using an arbitrary browser.
%\gls{oida} is written in Java and freely available as \gls{opensource}~\cite{openidattacker} (see also Section \ref{sec:implementation}).
We varied several parameters selected according to our white-box analysis, until the attack  was working.

%-----------------------
\paragraph{Exploit}
Finally, we performed the attacks in the web attacker model:
For only one attack (\gls{trc}) it is necessary that victim $\V$ visits a web page $SP_\A$ under control of the attacker $\A$.
In our setting, $\V$ is already authenticated to the trusted \gls{idp} (stored in a session cookie), so that no explicit authentication of $\V$ is necessary.
We verify that the token $t$ is indeed transferred to $SP_\A$, and that we could use this token from our second browser to gain access to the target $SP$.

To verify \gls{kc} attacks, we have sketched two strategies in \autoref{sec:attackclasses}.
For following the first strategy, the precondition that an association $\alpha$ exists between the target SP and the trusted \gls{idp} must be fulfilled.
We can get the value of $\alpha$ in message (5.) of  \autoref{fig:openid_simple} when we try to log in with the victim's identity.
This attempt will not succeed, but we can see message (5.) nonetheless.
We then established a new association between the target \gls{sp} and \gls{oida} using the same $\alpha$ and analyzed whether the target \gls{sp} afterwards accepted our malicious tokens as valid for $\V$.
For the second strategy, only an association $\beta$ between the target \gls{sp} and the malicious \gls{idp} is necessary.
We verified that the \gls{sp} accepted tokens containing $(\URLID_\V, \URLIdP_\V)$ that were signed with the malicious association $\beta$.

For the two remaining attacks (\gls{ids}, \gls{ds}), we only needed to know the first and second identity of $\V$. 
We verified that the target \gls{sp} accepted our malicious tokens for these two identities.

%We used the \gls{sp}'s login form and entered the attacker's identity $\URLID_\A$ so that the \gls{sp} will start the discovery and the \gls{asso} against our tool.
%Then, we configured it to apply the \emph{\acrlong{trc}}, \emph{\acrlong{kc}}, \emph{\acrlong{ids}} and \emph{\acrlong{ds}} attacks.

\section{Practical Evaluation}
\label{sec:evaluation}

\begin{overview}
    \begin{itemize}
       \item Erst ganz kurz (plus eine Tabelle) die Übersicht über alles was wir gemacht haben.
            \begin{itemize}
                \item Da sehe ich z.B.\ unsere Attack Scenarios vs implementations.
            \end{itemize}
        \item Dann im Detail einige ausgesuchte implementations/Webseiten. Da sehe ich Drupal/Sourceforge/Owncloud.
        \item Insgesamt so aufziehen, dass wir uns mehr auf implementations/Libraries beziehen.
            \begin{itemize}
                \item Viele Webseiten wie z.B.\ Wordpress nutzen janrain implementations
                    \begin{itemize}
                        \item mehr Beispiele hierfür finden
                    \end{itemize}
                \item Einige wenige haben eigene Implementierungen
                    \begin{itemize}
                        \item Owncloud, Drupal, Sourceforge
                    \end{itemize}
            \end{itemize}
\end{itemize}
\end{overview}

%According to the responsible disclosure model, w
We reported all vulnerabilities  to the liable security teams and to the Computer Emergency Response Team (CERT).
In case we got a response from the developers, the time to fix the reported issues ranged between a few days and several months. Furthermore, we supported the developer teams during fixing the reported issues.

Our results are summarized in \autoref{tab:results}: for \evalnumber{11} out of \evalnumber{16} targets, we were able to access a protected resource. On eight of the eleven targets an attacker can compromise \emph{all} of the accounts, without any user interaction. On the other three targets the account of any victim can be compromised, if he visits a malicious website.

 % to which he has no rights.

%---------------------------
\subsection{\acrlong{trc}}
\label{sec:trc_evaluation}

%In our evaluation we concentrated on \gls{oid} and
We analyzed the processing of the $\URLSP$-parameter.
To verify this vulnerability, we configured our custom \gls{idp} to create a token containing $\URLSP\_\A$ instead of the correct $\URLSP$.
If the token was accepted, we categorized the \gls{trc} attack scenario as applicable:
%The result of the evaluation can be seen in \autoref{tab:results}:
\evalnumber{6} out of \evalnumber{16} \gls{oid} targets were susceptible to the described \gls{trc} attack. %, i.e.\ the \gls{sp} does not validate $\URLSP$ parameter correctly.
%This allows unauthorized access by letting the victim click on a link.

\paragraph{JOID}
JOID~\cite{joid,joidack} is a free \gls{opensource} library supporting \gls{oid} authentication. At first, we evaluated whether the library verifies the $\URLSP$.
For that purpose, we used the \gls{oida} and configured it to create a token containing $\URLA$ instead of the original $\URLSP_{\text{JOID}}$.
Since the JOID \gls{sp} running on $\URLSP_{\text{JOID}}$ accepted the token, we started the second step of the analysis -- the exploit:

\begin{enumerate}
\item In the role of the attacker, we upload a website containing a \gls{php} script, which is available from the Internet by visiting $\URLA$, see \autoref{fig:attack_returnto}.
 \item We then simulate the victim who visits $\URLA$ from a different PC\@.
     For testing, we used a Yahoo account.
     The attacker's \gls{php} script generates a \emph{Token Request} for the victim containing $\URLSP=\URLA$ and then redirects him to $\URLIdP_\text{Yahoo}$.
 \item Still in the role of the victim, we do not need to authenticate to the \gls{idp} according to the methodology, see \autoref{sec:methodology}. Afterwards, the \gls{idp} generates the \gls{oid} token $t=(\URLID_\V,\URLA,\dots)$.
     Then, the \gls{idp} redirects the victim together with $t$ to the \gls{sp}, using $\URLA$ from the \emph{Token Request} -- the victim sends $t$ to the attacker's script.
 \item Once the script receives $t$, the attacker sends $t$ to the inspected JOID \gls{sp}. 
     Although $\URLA\neq \URLSP_{\text{JOID}}$, JOID accepts $t$ and the attacker is logged in with victim's account.
\end{enumerate}
\hspace{1pt}

%We contacted the developers and in collaboration, we created a patch fixing the issues\footnote{JOID was also vulnerable to \gls{ids} which will not be explained in detail here due to missing space}.
%They acknowledged our work in~\cite{joidack}.

%---------------------------------
\subsection{\acrlong{kc}}
\label{sec:keyconfusion_evaluation}
\evalnumber{Three} targets were vulnerable to \glsfirst{kc}: Drupal, Zend Framework and Sourceforge.
These implementations used a key belonging to \gls{oida} for verifying the signature instead of using the key belonging to the victim's \gls{idp}.
The attack on Drupal worked as follows:

%In this section, we describe the attack applied on Drupal.

\paragraph{Drupal}
\label{sec:drupal}
Drupal~\cite{drupal} is a free \gls{opensource} \gls{cms}.
It is based on \gls{php} and according to~\cite{w3techs}, it is the third most frequently used \gls{cms}.
Famous sites using Drupal are Twitter (Alexa rank 11) or Typepad (Alexa rank 498).
Its \gls{oid} support is shipped with every Drupal distribution and just needs to be activated within the settings menu.

%The investigation of Drupal's \gls{oid} implementation revealed a new attack class (\gls{kc}).

We started to analyze the implementation by carrying out the \gls{trc}, but it failed.
%Looking into the \gls{php} code revealed, that the implementation correctly verifies the $\URLSP$ parameter so that \gls{trc} was not possible.
%
Then, we tried to apply the \gls{ids} attack:
We submitted $\URLID_\A$ on the Drupal login form.
The \gls{sp} starts the discovery on it and receives $\URLIdP_\A$ belonging to our \gls{oida} \gls{idp}.
Drupal redirects us to it, but instead of creating a token for $\URLID_\A$, it creates a token $t^*=(\URLID_\V,\dots)$ containing the victim's Google identity.
Sending $t^*$ to Drupal did not succeed.
Drupal noticed that the originally submitted identity $\URLID_\A$ differs from the value $\URLID_\V$ contained in $t^*$.
As a result, Drupal starts a second discovery on $\URLID_\V$, which returns $\URLIdP_\V$. 
Drupal compares this value to $\URLIdP_\A$ returned by the first discovery.
Since the values are not equal, we are not logged in.
Interestingly, Drupal does \emph{not} compare the discovered value with the value $\URLIdP$ contained in $t^*$, thus sending a token $t^*=(\URLID_\V,\URLIdP_\V,\dots)$ also fails.

In order to prevent the second discovery process, which mitigates the attack, we did a white-box analysis of the source code. We found out that Drupal uses the \gls{php} \SESSION{} variable to store and load $\URLID$ and $\URLIdP$. In this manner, Drupal links both messages: the login request and the received token.

The \SESSION{} variable is a globally available \gls{php} array which holds arbitrary session data on a per-user basis.
Whenever Drupal receives an \gls{oid} token $t^*$, it first verifies if the $\URLID$ parameter, contained in $t^*$, matches the value stored in \SESSION{}.
If they differ, as in the case of the \gls{ids} attack, Drupal starts again a discovery on $\URLID$ contained in $t^*$.
The discovery returns the corresponding $\URLIdP$ and if these values do not match the $\URLIdP$ parameter stored in \SESSION, $t^*$ is not accepted.

To finally prevent the second discovery and to bypass the verification logic, we had to overwrite the \SESSION{} variable.
The attack is shown in \autoref{fig:attack_drupal} and works as follows:

\begin{figure*}[!ht]
    \centering
    \includegraphics[width=0.9\linewidth]{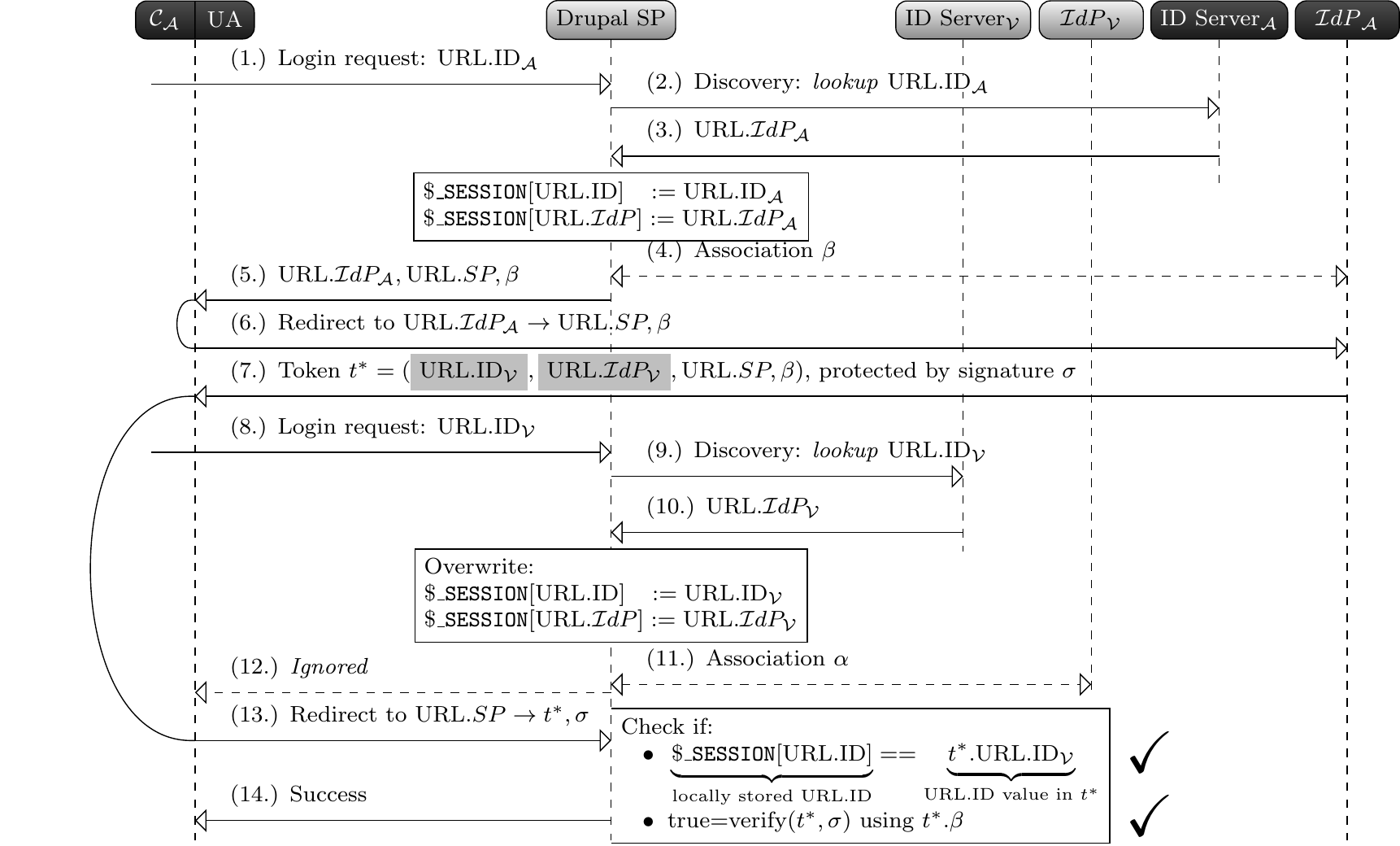}
    \caption{\acrlong{kc} attack on Drupal: Before the token $t^*$ in Step (7.) is forwarded to Drupal in Step (13.), the attacker $\C_\A$ starts a second login request in Step (8.) using the victim's identity $\URLID_\V$.
        This overwrites the $\URLID$ and $\URLIdP$ data stored in \SESSION{} and prevents the second discovery.}
    \label{fig:attack_drupal}
\end{figure*}

\begin{itemize}
    \item[(1.)-(3.)] A login request with the attacker's account $\URLID_\A$ is started.
        Drupal discovers it and stores $\URLID_\A$ and $\URLIdP_\A$ in \SESSION{}.
    \item[(4.)] Drupal starts an \gls{asso} with $\IdPA$, which returns $\beta$ (using \gls{kc} strategy 2).
    \item[(5.)-(7.)] Drupal redirects the attacker to $\URLIdP_\A$.
        The \gls{oida} \gls{idp} creates a token $t^*=(\URLID_\V,$ $\URLIdP_\V,\URLSP,\beta)$.
        Then, the attacker delays the sending of the token to Drupal.
    \item[(8.)-(10.)] The attacker submits a further login request to Drupal, but this time with the victim's identity $\URLID_\V$.
        Drupal starts a new discovery on it and receives $\URLIdP_\V$.
        Both values, $\URLID_\V$ and $\URLIdP_\V$, are then stored in \SESSION{}, overwriting $\URLID_\A$ and $\URLIdP_\A$.
    \item[(11.)] Drupal starts another \gls{asso} with $\IdPV$, which returns $\alpha$.
    \item[(12.)] Drupal redirects the attacker to $\URLIdP_\V$, but this redirect is not relevant for the attack.
    \item[(13.)-(14.)] The halted token $t$ in (6.) is now sent to Drupal.
Drupal verifies the signature.
The interesting point at this step is that Drupal loaded the key from the database by only using $\beta$ contained in $t^*$.
It does not verify whether the \gls{asso} $\beta$ was really established with $\URLIdP_\V$.
Thus, the signature is valid.
Then, Drupal compares the values of $\URLID_\V$ and $\URLIdP_\V$ contained in the token with the ones stored in \SESSION{}.
Because of being equal, there is no second discovery and we are logged in with the victim's identity.
\end{itemize}
\hspace{1pt}
We reported the issue to the Drupal security team and suggested to fix it by fetching the key via $(\URLIdP,\alpha/\beta)$ instead of using $\alpha$/$\beta$ only.
They accepted the idea and implemented it in their new release Drupal 8 as well as in Drupal 7 and Drupal 6~\cite{CVE-2014-1475}.
For a better understanding, we added a video as a demonstration of this attack that shows the usage of \gls{oida}~\cite{attackingdrupal}.

%-----------------------------
\subsection{\acrlong{ids}}
\label{sec:idspoofing_evaluation}

\evalnumber{Six} of the tested targets were vulnerable to \gls{ids}.
Those targets did not check if the identity contained in the token was issued by the correct \gls{idp}. % and we could impersonate the victim's Google or Yahoo account without his interaction.
%As a result, we were logged in at the \gls{sp} with the victim's Google and Yahoo accounts without knowing the corresponding password.

\paragraph{Sourceforge}
\label{sec:sourceforge}
\begin{overview}
\begin{itemize}
    \item Wir konnten hier einfaches \gls{ids} durchführen
    \item Nachdem wir die Betreiber der Seite kontaktiert haben, wurde der Bug innerhalb weniger Tage (8-9) Behoben.
    \item Jedoch war den extended \gls{ids} (über \SESSION{} möglich)
    \item Dieser wurde nach 3 Tagen gefixt.
\end{itemize}
\end{overview}

\begin{figure}[h]
    \centering
    \includegraphics[width=\linewidth]{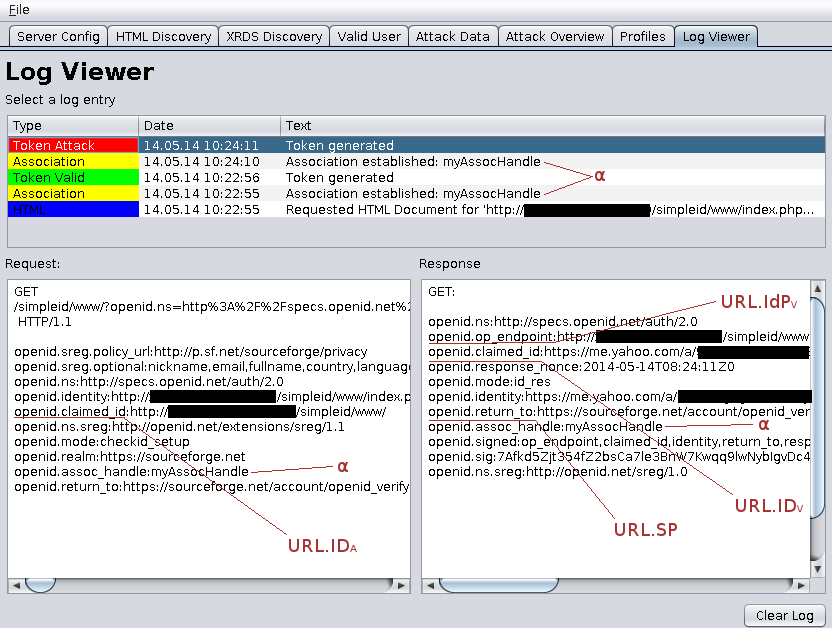}
    \caption{\gls{ids} attack on Sourceforge. The \gls{oida} \emph{log viewer} window lists all exchanged \gls{oid} messages. The Screenshot shows that the \gls{sp} requests a token for $\URLID_\A$, but the tools ignores the wish and responds with a token for $\URLID_\V$.}
    \label{fig:openid_attacker_log}
\end{figure}

%Sourceforge~\cite{sourceforge} is a widely used web-based source code repository (Alexa Rank 160) and it allows its users to connect an \gls{oid} to their account.
Initially, we started a black-box testing and detected that the applied \gls{oid} authentication is vulnerable against \gls{ids}. 
\autoref{fig:openid_attacker_log} shows the log window of our developed \gls{oida} tool which contains all exchanged \gls{oid} parameters.
Consequentially, we contacted the support team and described the issue. Later on, they answered us that vulnerability is fixed.

We analyzed the implementation again. Using \gls{oida}, we found out that the \gls{ids} attack was no longer working. Unfortunately, Sourceforge noticed that the identity contained in the \gls{oid} token is different to the one requested during the login request. As a result, Sourceforge started to rediscover the submitted identity and the attack failed. Based on this information, we mounted the same attack technique as on Drupal: we confused Sourceforge with a second login request on $\URLID_\V$ right after we submitted the login request for $\URLID_\A$. As expected, \gls{kc} was applicable. Consequentially, we contacted the Sourceforge support team and described our finding.

Using only black-box testing we were able to determine that the implementation uses a session variable to connect the initial login request with the token response.
In collaboration with the support team, we fixed the vulnerability.
We suggested to fetch the key from the database not only by using $\alpha$, but rather by a combination of $(\URLIdP,\alpha)$.
In this manner, \gls{kc} can be prevented.
The Sourceforge support team pointed out that the \gls{oid} specification~\cite[Section 11.2]{openid20} addresses this problem, but it only describes that a rediscovery is necessary in the given case.
It neither addresses how to find out that the identity of the login request is not the same as in the token, nor mentions that this fact could be abused by attackers.

%--------------------------
\subsection{\acrlong{ds}}
\label{sec:discoverysoofing_evaluation}

\Gls{oc} is up to now the only framework that is vulnerable to  \gls{ds} attacks.
We nevertheless describe this attack because it allows us to utilize the discovery phase for the injection of identities, which are not controlled by our \gls{idp}, but used for the login.

\begin{overview}
    \begin{itemize}
        \item Frage: Soll das extendend \gls{ids} erst hier beschrieben werden?
        \item Oder bereits zuvor?
        \item OpenID support removed?
    \end{itemize}
\end{overview}

\paragraph{\glsentrytext{oc}}
\label{sec:owncloud}
\begin{overview}
    \begin{itemize}
        \item Owncloud erlaubt nur OpenID 1.0
        \item Neuer Angriff entdeckt: \gls{ds}
            \begin{itemize}
                \item Innerhalb der HTML Datei wurde eine fremde Identity eingetragen (z.B. google).
                \item Der vom IdP zum SP gesendete Token selbst enthält aber die Identität der Angreifers, nicht die des Opfers.
                \item Der SP führt aber auf der Identität erneut ein Discovery durch und verwendet die darin enthaltene fremde Identity.
            \end{itemize}
        \item Behebung dauerte fast einen Monat
        \item OpenID support removed?
    \end{itemize}
\end{overview}
\Gls{oc}~\cite{owncloud} is a \gls{php}-based, \gls{opensource} cloud framework.
It provides universal access to files as a \emph{self}-controlled alternative to Dropbox or Google Drive and additionally,
\gls{oc} users' can sync private data such as contacts and calendar information.
\gls{oc} allows \gls{sso} by simply activating the \gls{oid} plugin, which is distributed by default with \gls{oc} 5.

%While investigating \gls{oc}'s \gls{oid} implementation, we discovered a new attack technique (\gls{ds}).

% To understand this technique, one has to understand how \gls{oc} processes \gls{oid}: \\

% \begin{compactenum}
    % \item At first, the client submits $\URLID_\C$ via the login-form.
    % \item \Gls{oc} then fetches the HTML document at $\URLID_\C$.
    % \item \Gls{oc} extracts the $\URLIdP_\C$ contained in it.
        % % Optionally, it is possible to include an $\URLID$ as well.
    % \item \Gls{oc} creates the authentication request message and sends it back to the client, redirecting him to $\URLIdP_\C$.
   % \item The \gls{idp} creates the token $t$ as usual and redirects the client back to $\URLSP$, i.e.\ back to \gls{oc}.
   % \item \Gls{oc} performs a rediscovery on $\URLID_\C$ contained in $t$ as described above in 2.
   % \item \Gls{oc} itself does not verify the token's signature -- it sends the token back to the \gls{idp} which verifies it as shown in Step (10.) and (11.) of \autoref{fig:openid_simple}.
      % It uses the \gls{idp} URL that was extracted in the rediscovery.
  % \item If the \gls{idp} returns that the token is valid, the user is logged in.
% \end{compactenum}

There are two interesting parts in \gls{oc}'s \gls{oid} implementation:
\begin{inparaenum}
    \item in comparison to other implementations, \gls{oc} always starts a rediscovery so that the \gls{kc} attack is not applicable.
    \item \gls{oc} does not verify the token's signature itself. Instead, it uses the \emph{check authentication} mechanism (see steps (12.) and (13.) in \autoref{fig:openid_simple}) and sends the token to the \gls{idp}.
\end{inparaenum}
This means that using \gls{oida} to send, for example, a token for a Google account would lead \gls{oc} to send the token directly to a Google server for verification, which will not accept it.
Thus, for attacking \gls{oc} we could not send a token containing $\URLID_\V$ -- the \gls{ids} and \gls{kc} attacks are not possible.

% One of the ideas behind \gls{oid} is to separate the user's login URL from its concrete \gls{idp}
% as shown in \autoref{fig:openid_simple}.
% For example a user's identity URL could by \url{http://mysite.com/myid}.
% On this URL, the user can place an \gls{xrds} document which can define an arbitrary \gls{idp}, for example, Google, Yahoo, etc.
% This advantage of this approach is, that the user can easily change one \gls{idp} to another one, because the \gls{sp} only uses \url{http://mysite.com/myid} to log in the user, which is independent of concrete \gls{idp}.

\begin{figure}[!ht]
    \centering
    \includegraphics[width=0.9\linewidth]{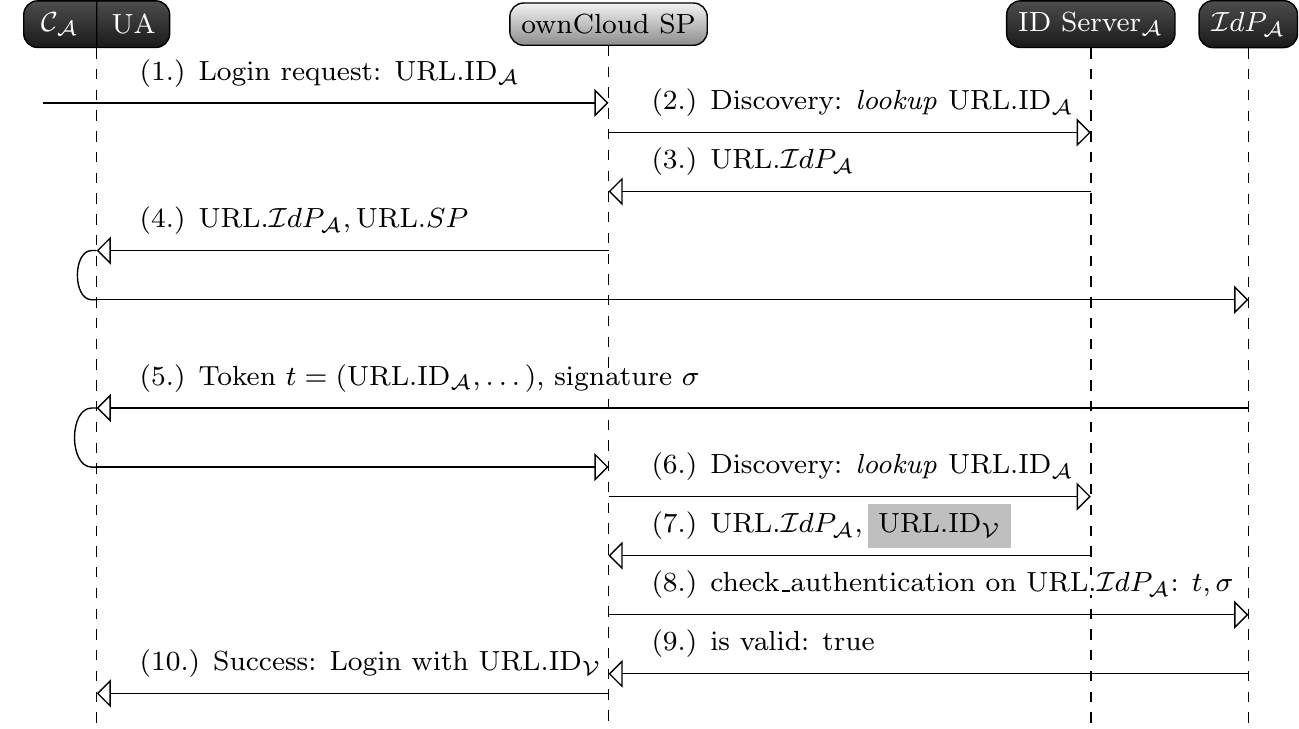}
    \caption{The \acrlong{ds} attack on \glsentrytext{oc}: The attacker's ID server returns $\URLID_\V$ on the second discovery.
             \gls{oc} uses this identity value for the login instead of the identity provided within the token.}
    \label{fig:attack_owncloud}
\end{figure}

By examining the \gls{oid}'s discovery phase, we found out that the \gls{oid} specification allows the usage of an $\URLID$ value in the HTML/XRDS files.
This feature can be used to trick \gls{oc} as shown in \autoref{fig:attack_owncloud}.

When \gls{oc} receives the \gls{oid} token in Step (5.), it performs a rediscovery on the contained identity.
We configured the \gls{oida} \gls{idp} to include the victim's identity $\URLID_\V$ in the discovered document of Step (7.) as shown  in \autoref{lst:html_discovery_id} additionally to $\URLIdP_\A$.
Afterwards, \gls{oc} sends the token to the attacker's $\IdPA$ in Step (8.) by using the discovered $\URLIdP_\A$ and it returns that the token is valid in Step (9.).
Surprisingly, instead of using the $\URLID_\A$ contained in $t$ to log in the user, \gls{oc} uses $\URLID_\V$ returned in Step (7.).
We were logged in with the victim's identity.

We contacted the \gls{oc} security team and reported the issue.
The \gls{oc} team acknowledged our work in~\cite{owncloudack}.
% As a result, they decided to remove the \gls{oid} module from their next major version \gls{oc} 6 due to lack of time shortly before its release and acknowledged our work in~\cite{CVE-2014-2048}.

\subsection{Additional Findings}
\label{sec:additionalfindings}
%During the research, we provided further investigations regarding the authentication mechanisms. 
%These investigations are related to the security of the application, but do not result in unauthorized access.
%The first two tests are related to the \emph{Message Parsing} logic, the last test is based on the examination of the token's \emph{Freshness} described in \autoref{sec:analyzingmodel}. 

The findings described here did not result in a valid attack according to our model, but are worth reporting.

%\paragraph{Standard Conformity}
%\label{sec:standardConformity}
%% According to the \gls{oid} specification~\cite[Section 10.1]{openid20}, the following parameters must be signed: \texttt{op\_endpoint}, \texttt{return\_to}, \texttt{response\_nonce}, \texttt{assoc\_handle}, \texttt{claimed\_id} and \texttt{identity}. 
%The \gls{oid} specification~\cite[Section 10.1]{openid20} claims to force the signing of important parameters, for example, $\URLID$, $\URLIdP$, $\URLSP$, \dots
%During our analysis, we evaluated whether the \gls{sp} verifies if each of the parameter mentioned by the specification is signed.
%
%The test consists of two parts:
%\begin{inparaenum}
%    \item presence of the signature parameter at all and its validation plus
%    \item the completeness parameters claimed to be signed.
%\end{inparaenum}
%
%According to the first part, we found one implementation (CFOpenID), which did not validate the signature at all. 
%In this manner, no integrity protection is provided and every token is valid.
%In the second test, \evalnumber{4} of \evalnumber{16} targets (CFOpenID, OpenID CFC, OpenID 4 Node.js, Zend Framework) accepted tokens, in which only parts of all required parameters are signed.
%In fact, they accepted a token with at least one arbitrary signed parameter.
%This finding can be crucial regarding the security, if an attacker gets access to such tokens. 
%Consequentially, he can adapt all unsigned parameters and impersonate other users on the \gls{sp}.

\paragraph{Unsigned OpenID Parameters}
The \gls{oid} specification~\cite[Section 10.1]{openid20} requires the following parameters to be signed: \texttt{op\_endpoint}, \texttt{return\_to}, \newline \texttt{response\_nonce}, \texttt{assoc\_handle}, \texttt{claimed\_id} and \texttt{identity}.  \evalnumber{4} of \evalnumber{16} targets (CFOpenID, OpenID CFC, OpenID 4 Node.js, Zend Framework) accept tokens in which some of these parameters were not signed, and could thus be forged by an attacker.

\paragraph{XML External Entity}
\label{sec:xxe}
% During the \gls{oid} discovery phase, the \gls{sp} processes HTML or XRDS documents. 
% Since XRDS documents are XML-based,
%An attacker can exploit vulnerabilities in the XML processing logic.
%For instance, XML-based services are endangered against efficient \gls{dos} attacks~\cite{JS_WS_DoS2013} or External Entity attacks. 
%Thus, every implementation working with XML documents should provide countermeasures mitigating these attacks.
%With respect to \gls{oid}, the processing of \gls{xrds} documents during the discovery phase can be a target of such attacks~\cite{Facebook:Silva:XXE:2014} allowing to read arbitrary files on the hosting server.
%As a result of our evaluation, w
We determined that \evalnumber{2} of \evalnumber{16} analyzed targets (OpenID CFC, Net::OpenID::Consumer) are susceptible to XXE attacks~\cite{XXE_OWASP,Facebook:Silva:XXE:2014}.
Additionally, we found out that Slashdot~\cite{slashdot} (Alexa rank 1626) was vulnerable to XXE because of using the Net::OpenID::Consumer library. Slashdot acknowledged our findings in~\cite{slashdotAck}
%We reported it to their support team and afterwards, they disabled the \gls{oid} login.

\paragraph{Replay Attack}
\label{sec:replayattacks}
%Every \gls{sso} protocol provides parameters restricting the reuse of an authentication token and its lifetime to prevent replay attacks.
%% \gls{oid} proves the freshness of the token via the \texttt{openid.response\_nonce} parameter. 
%According to the protocol specification~\cite[Section 11.3]{openid20}, the prohibition of reuse is optional and is not categorized as critical.
%
%With respect to security, the validation of the timestamps is more significant. 
%If this check is not correctly performed, the token's lifetime is infinite. 
\gls{oid} has only one parameter containing a timestamp (\texttt{openid.\ response\_nonce}).
It contains the creation time of the token concatenated with a random string, but does not include an expiration time.
Thus, the \gls{sp} can decide on its own how long it accepts such a token.
%Because of belonging to the list of parameters that must be signed, this parameter cannot be changed by an attacker. 

The lifetime of a token is additionally limited by the lifetime of the \gls{asso} and the corresponding key.
%This lifetime is defined by the \gls{idp} during the \gls{asso} phase.
We found that this lifetime varies heavily: \glspl{asso} with Yahoo have a lifetime of 4 hours, with Google 13 hours, and with MyOpenID 14 days.
%In conclusion, the lifetime of \gls{oid} tokens is usually not very long, which makes replay attacks hard to apply.

%During our research, we analyzed the \glspl{sp}.  \todo{The results are ...}

% \subsection{Key Confusion}
% An applicable attack scenario in this step was provided by Wang et al.~\cite{microsoft}.
% The authors described an attack which enforces the generation of an authentication token containing unsigned required attributes, but exactly those attributes were used for the authentication.
% Since the injection of malicious content and the manipulation of these attributes was applicable, the impersonation of other users was possible.
% Another attack technique is ``Signature exclusion'' relying on poor implementation of a server's security logic~\cite{somorovskySAML}.
% In this scenario the verification logic proves the validity of a signature only if a signature is provided.
% If no signature is present, the validation is skipped.

\section{OpenID Attacker \\ Implementing a malicious \gls{idp}}
\label{sec:implementation}

\begin{overview}
\todo{Christian}
\begin{itemize}
    \item Wichtig: Attacker IdP kann valide Token ausstellen.
    \item Und: Der Attacker IdP kann zusätzlich Token für andere IdPs (z.B.\ Google) ausstellen
\end{itemize}
\end{overview}

\begin{figure}[h]
    \centering
    \includegraphics[width=\linewidth]{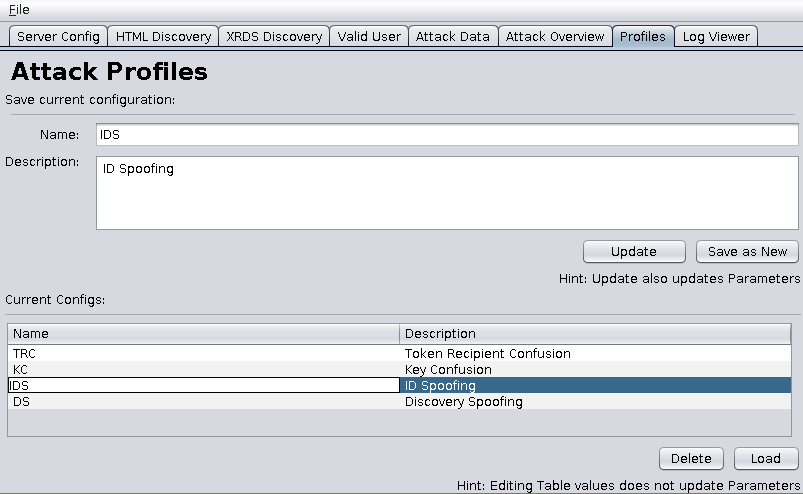}
    \caption{The \gls{oida} profile window allows to automatically chose an attack configuration for all four presented attacks.
A video as a demonstration of the attack on Drupal showing the usage of \gls{oida} can be found on~\cite{attackingdrupal}.
    }
    \label{fig:openid_attacker_profiles}
\end{figure}

We developed \emph{\gls{oida}} as a part of our research and as a result of our token verification model for \glspl{sp}.
\gls{oida} is an \gls{opensource} \gls{pentest} tool that mainly acts as an \gls{oid} \gls{idp} and offers a \gls{gui} for easy configuration, see Figures \ref{fig:openid_attacker_log} and \ref{fig:openid_attacker_profiles}.
As such, it is able to operate during all three phases of the \gls{oid} \gls{sso} protocol.
\gls{oida} is free, \gls{opensource} and can be downloaded here~\cite{openidattacker}.

\begin{figure}[!ht]
    \centering
    \includegraphics[width=0.9\linewidth]{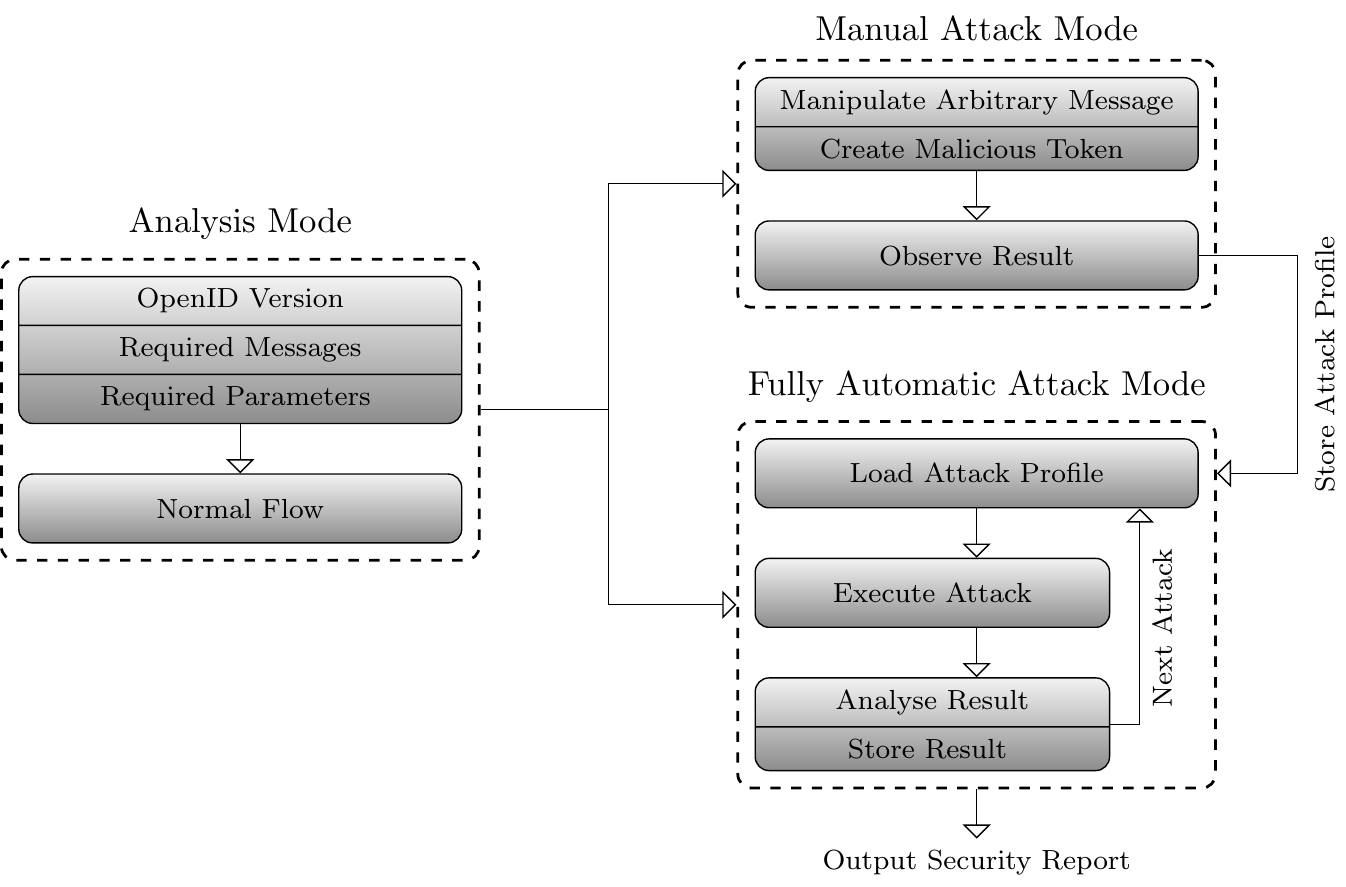}
    \caption{The three modes of \gls{oida}.}
    \label{fig:openid_attacker_modes}
\end{figure}
The main advantage of \gls{oida} is its flexibility -- the attacks can be provided manually or full automatically. As shown in \autoref{fig:openid_attacker_modes}, \gls{oida} works in three modes:
\begin{inparaenum}
\item Analysis, 
\item Manual Attack,
\item Fully Automatic Attack.
\end{inparaenum}

\paragraph{Analysis Mode}
In this mode \gls{oida} is used to analyze the normal behavior of the target \gls{sp}. 
For this purpose, \gls{oida} acts as benign \gls{idp} and creates valid tokens. 
Additionally, it gets and stores information about the supported \gls{oid} version on the \gls{sp}, the exact message flow, for instance, which optional messages are used, the schema of the messages and the required parameters. 
Once calibrated, \gls{oida} stores the collected informations in a data structure, called \emph{Normal Flow}, used by the other modes.

In this mode \gls{oida} works automatically and does not require any interaction or configuration.

\paragraph{Manual Attack Mode}
In this mode, \gls{oida} acts as a malicious \gls{idp} hand-operated by the attacker. 
The attacker starts the security analysis on basis of the informations stored in \emph{Normal Flow}. He manipulates parameters in the messages and creates malicious tokens. He then observes the results of the attacks. 
In this mode, the attacks and the evaluation of the attacks are carried out manually.

The idea behind the \emph{Manual Mode} is the fact that new attack vectors can be inspected. 
This is an important fact, because the \emph{Manual Mode} allows to investigate the \gls{oid} protocol very deeply and fine granular as every single aspect of the protocol can be manipulated.
In combination with a running \gls{sp} implementation in debugging mode, this mode helps to understand the source code of the \gls{sp} to find implementation as well as protocol issues.
We used this mode to discover the four novel \gls{oid} attacks \gls{trc}, \gls{ids}, \gls{kc}, \gls{ds} during a white-box analysis.
The configuration of the attack vectors can then be stored as an \emph{Attack Profiles} (cf. \autoref{fig:openid_attacker_profiles}) and can later be loaded for black-box analysis. 
By using \emph{Attack Profiles}, any of the stored attacks can be easily reproduced with only one click.

\begin{figure*}[!ht]
    \centering
    \includegraphics[width=0.9\linewidth]{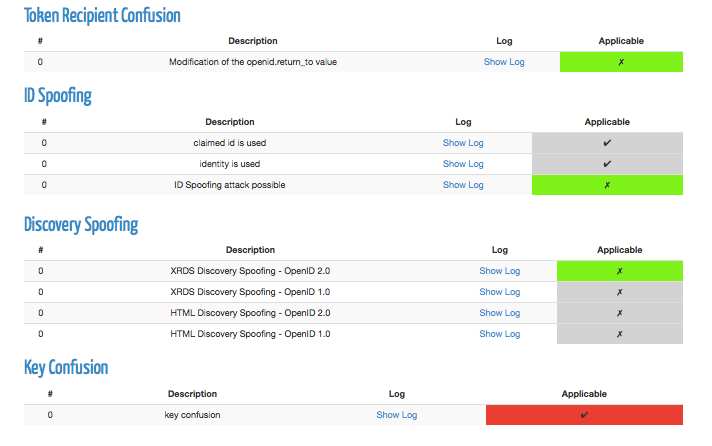}
    \caption{The \emph{Fully Automatic Attack Mode} outputs a security report.}
    \label{fig:openid_attacker_report_2_extracted}
\end{figure*}

\paragraph{Full Automatic Attack Mode}
In this mode \gls{oida} acts as full automated malicious \gls{idp} \gls{pentest} tool. Initial, \gls{oida} loads the stored \emph{Attack Profiles} and the \emph{Normal Flow}.
Afterwards, it sequentially executes the attacks defined in the profiles. 
Then, \gls{oida} analyzes the result of the attack and stores the information in the security report, see \autoref{fig:openid_attacker_report_2_extracted}. 
In conclusion, \gls{oida} summarizes the results of all attacks contained in the \emph{Attack Profiles} and creates a security report.

\section{Related Work}
\label{sec:relatedwork}

Related work can be divided into three parts:
research on analysis of \gls{sso} systems,
specific investigations in the field of \gls{oid}, and
development of SSO testing tools.
Please note that none of the previous papers considers malicious \glspl{idp} as part of the attacker, and none of the \gls{oid} papers considered attacks on the association phase.

\begin{overview}
Protocol Analysis
\begin{itemize}
    \item BAN-logic, \cite{banLogic}
    \item \cite{millen1995interrogator}
    \item \cite{nrlLanguageGeneration}
        \begin{itemize}
            \item NRL Protocol Analyzer is a tool for proving security properties of cryptographic protocols
            \item Can find flaws if they exist.
            \item In its very basic, it is a search tool
            \item Paper presents a procedure for automatic language generation in the NRL Protocol Analyzer, which is a hard work otherwise.
        \end{itemize}
\end{itemize}
\end{overview}

%Initially, in 1989 Burrows et al.~\cite{banLogic} created the BAN-logic, a formal analysis system for cryptographic protocols.
%Later on, software tools for protocol verification were created and used~\cite{millen1995interrogator, nrlLanguageGeneration}.

\begin{overview}
\gls{saml} (vermutlich sollten wir nicht zuviel auf \gls{saml} rumreiten)
\begin{itemize}
    \item \cite{gross\gls{saml}}
        \begin{itemize}
            \item Security Analysis of the \gls{saml} Single Sign-on Browser/Artifact Profile
            \item Connection hijacking / replay attack
            \item man-in-the-middle attacks
            \item HTTP referrer attack
        \end{itemize}
    \item \cite{somorovsky\gls{saml}}
        \begin{itemize}
            \item Abuses features of XML to perform XML Signature Wrapping attacks
            \item Breaks 11 out of 14 \gls{saml} frameworks.
        \end{itemize}
    \item \cite{sun2012devil}
\end{itemize}
\end{overview}

\paragraph{\gls{sso} Security}
Various vulnerabilities have been found over the last two decades.
In 2003 and 2006, Groß~\cite{grossSAML,GP06SAML} analyzed the \gls{saml} Browser/Artifact profile and identified several flaws in the \gls{saml} specification that allow connection hijacking/replay attacks, as well as \gls{mitm} attacks and HTTP referrer attacks. We used these attacks as model for the \gls{trc} attack.
In 2008 and 2011, Armando et al.~\cite{Armando.2008,Armando.2011} built a formal model of the \gls{saml} V2.0 Web Browser SSO protocol and analyzed it with the model checker SATMC.
The authors found vulnerabilities in Google's \gls{saml} interface.
% In 2008, Armando et al.~\cite{Armando.2008} built a formal model of the \gls{saml} V2.0 Web Browser SSO protocol and analyzed it with the model checker SATMC. As a result, the authors found a vulnerability in Google's \gls{saml} interface. Later on, same authors identified yet another attack on the \gls{saml}-based SSO of Google Apps~\cite{Armando.2011}.
In 2012, Somorovsky et al.~\cite{somorovskySAML} investigated the \gls{xmlsig} validation of several \gls{saml} frameworks.
By using the \gls{xsw} attack technique,
% originally published by McIntosh and Austel~\cite{McIntosh2005} in 2005,
they bypassed the authentication mechanism in 11 out of 14 \gls{saml} frameworks.

Sun et al.~\cite{sun2012devil} analyzed the implementation of nearly 100 \gls{oauth} implementations, and found serious security flaws in many of them.
Their research concentrated on classical web attacks like \gls{xss}, \gls{csrf} and TLS misconfigurations. Further security flaws in \gls{oauth} based applications were discovered by \cite{HomakovGitHubHackOAuth,HomakovFacebookHackOAuth,HomakovOAuthQuestions,GoldschlagerFacebookHack,GoldschlagerFacebookHackAgain,ssoscan}, whereby the authors concentrated on individual attacks. 
In 2013 Wang et al. introduced a systematic process for identifying critical assumptions in SDKs, which led to the identification of exploits in constructed apps resulting in changes in the OAuth 2.0 specification~\cite{Wang:2013:ESU:2534766.2534801}.
Chen et al. revealed in 2014 serious vulnerabilities in \gls{oauth} applications on mobile devices caused by the developer's misinterpretation of the \gls{oauth} protocol~\cite{demystifiedOAuthCCS14}.

In 2014 Fett et al.~\cite{FettKuestersSchmitz-SP-2014} built a formal model of the BrowserID protocol~\cite{browserID}, which allows them to remodel known weaknesses and vulnerabilities in BrowserID.

\begin{overview}
OpenID
\begin{itemize}
    \item \cite{tsyr}
        \begin{itemize}
            \item presented several OpenID 1.0 related attacks at Black Hat in...
            \item E.g.\ downloading a malicious OpenID endpoint URL like localhost files or a large movie from YouTube.
            \item Malicious SP can redirect the user to ca malicious IdP to steal his password.
            \item IdP can track users.
            \item Replay attacks by stealing redirection link.
            \item CSRF
        \end{itemize}
    \item \cite{sessionSwap}
        \begin{itemize}
            \item Session Swapping attack
            \item attacker gets logged in with victim's account on SP
            \item implies all messages are sent over HTTP
            \item Does not look at association
        \end{itemize}
    \item \cite{kohlarOpenid}
        \begin{itemize}
            \item showed in \dots how identity information set within OpenID messages could be manipulated if the verification logic is improper and the authentication logic is not integrity protected.
            \item Therefore, they appended additional SReg and Ax extension parameters to token, but does not add them to \texttt{openid.signed}.
            \item Parameter injection
            \item Parameter forgery
        \end{itemize}
    % \item \cite{miculan2011formal}
    \item \cite{microsoft}
        \begin{itemize}
            \item Concentrate on real life systems, no formal analysis
            \item black-box
            \item browser relayed messages (BRMs)
            \item Related Work Section am Ende.
        \end{itemize}
\end{itemize}
\end{overview}

\paragraph{\gls{oid} Security}
The analysis of the \gls{oid} protocol started with version 1.0.
Eugene Tsyrklevich and Vlad Tsyrklevich~\cite{tsyr} presented several attacks on this \gls{oid} version at Black Hat in 2007.
They identified, for instance, a threat in the \gls{idp} endpoint URL ($\URLIdP$) published within the discovery phase.
It can point to critical files on the local machine or can even be abused in order to start a \gls{dos} attack by enforcing the \gls{sp} to download a large movie file.
% Additionally, they showed the attack potential of a malicious \gls{sp} which redirects the user to an attacker controlled \gls{idp}, and also the potential of a malicious \gls{idp}, which can track its users.
Comparable to \cite{sun2012devil}, they also looked at replay and \gls{csrf} attacks.
In 2008, Newman and Lingamneni~\cite{sessionSwap} created a model checker for \gls{oid} 2.0, but for simplicity, they removed the \gls{asso} phase out of their model.
By using it, they could identify a session swapping vulnerability, which enforces the victim to log in into attacker's account on an \gls{sp}.
In this manner, an attacker could eavesdrop the victim's activities.
In comparison to our work, the attacks presented in~\cite{sessionSwap} do not result in unauthorized access.
Interestingly, the authors of the paper modeled an \gls{idp} capable to make associations with legitimate \glspl{sp}.
However, they did not consider a dishonest \gls{idp} capable to start attacks like \gls{ids} and \gls{ds}.
Since \gls{kc} is related to the \gls{asso} phase, the attack was not covered by the model checker.
Later on, Sun et al.~\cite{journals/compsec/SunHB12} provide a comprehensive formal analysis on \gls{oid} and an empirical evaluation of 132 popular websites.
The authors investigated CSRF, Man-in-the-middle attacks and the SSL support of \gls{oid} implementations.
In contrast to our work, they assumed that the \gls{sp} and the \gls{idp} were trustworthy, so that they could not identify any of the attacks presented in this paper.
% Sollen wir wirklich aus der SICHERHEIT2010 zitieren?
% Sovis et al.~\cite{kohlarOpenid} showed in 2010 how identity information, set within \gls{oid} messages, can be manipulated if the verification logic is improper and the authentication logic is not integrity protected.
% They showed a technique named parameter injection which could be used to append additional \gls{sreg} and \gls{ax} extension parameters without invalidating the token's signature.
% However, the problem with this attack is, that the additional parameters are not processed by most \glspl{sp}, because they are not requested.
% Thus they described the parameter forgery attack, which allows to manipulate and change the value of extension parameters which are requested and therefore processed by an \gls{sp}.

Finally, Wang et al.~\cite{microsoft} concentrated on real-life \gls{sso} systems instead of a formal analysis. 
They  have well demonstrated the problems related to token verification with different attacks.
They developed a tool named BRM-Analyzer that handles the \gls{sp} and \gls{idp} as black-boxes by analyzing only the traffic visible within the browser.
Their paper served as a model for our research.
% Although their approach can be adopted to arbitrary \gls{sso} systems, they mainly concentrate on \gls{oid}.
However, the BRM-Analyzer is rather passive (it analyzes the browser related messages), while \gls{oida} acts as an \gls{idp} and as such, it can actively interfere with the OpenID workflow (e.g.\ create \gls{sso} tokens).
% This restriction also includes that they do respect the threat of a malicious \gls{idp}, which the presented \gls{oida} is.

In 2014, Silva et al.~\cite{Facebook:Silva:XXE:2014} exploited an XML External Entity vulnerability in Facebook's parsing mechanism of \gls{xrds} documents during the discovery phase.
The same attack is supported by the \gls{oida} and is part of our evaluation.
Simultaneously to our research, in 2014 Wang et al.~\cite{Cnet:CoveredRedirect:OpenIDOAuth} reported serious flaws in OAuth and OpenID, which are related to \gls{trc}.

\begin{overview}
Other Tools
\begin{itemize}
    \item NoTamper~\cite{notamper}
        \begin{itemize}
            \item Not SSO specific.
            \item Parameter tampering in Web Applications.
            \item Identifies vulnerabilities of type: $p_{server} < p_{client}$
            \item Nett formuliert: \emph{Finally, due to the inherent limitations of black-box analysis, our approach cannot offer guarantees of completeness; rather, we justify the utility of our approach by the severity of the real vulnerabilities we have discovered.}
            \item Evaluation: Only 8 open source applications and 5 live websites.
            \item Related Work Section am Ende.
        \end{itemize}
    \item InteGuard~\cite{integuard}
        \begin{itemize}
            \item \emph{InteGuard, the first system that offers security protection to vulnerable web API integrations.}
            \item \emph{We plan to make available the source code of InteGuard
            \item \emph{InteGuard is just a first step toward secure web service integration, offering only limited protection: for example, it does not protect provider-side flaws.}
in the near future and further improve its design.}
            \item Related Work Section am Ende.
        \end{itemize}
    \item AuthScan~\cite{authscan}
        \begin{itemize}
            \item Supports multiple SSO protocols.
            \item Protocol specification read from JavaScript code.
            \item Only a few attacks supported: MitM, replay attack, guessable token
            \item Related Work Section am Ende.
        \end{itemize}
\end{itemize}
\end{overview}

% \todo{The next paragraph is moved from Analyzing model/Schema and must be adjusted.}
%
% There are many security researches and practical evaluations related to this verification step, introducing a variety of attacks classes.
% Many of these attacks revealed critical vulnerabilities breaching the security of the end-systems.
% For instance, XML-based services are endangered against efficient \gls{dos}~\cite{JS_WS_DoS2013} attacks or External Entity attacks~\cite{XXE_OWASP}.
% Thus, every implementation working with XML documents should provide countermeasures to mitigate these attacks.
% With respect to SSO, the processing of XML-based \gls{saml} tokens~\todo{CVE-2013-6440} or for \gls{oid}, the parsing of \gls{xrds} documents during the discovery phase can be a target of such attacks as a recent attack on Facebook revealed~\cite{Facebook:Silva:XXE:2014}.
% Aside from XML, in 2009 security researchers used a vulnerability in Flickr's API based on wrong parameter parsing~\cite{duong2009flickr}.
% This allowed the infiltration of false content in spite of the provided integrity protection.

\paragraph{\gls{sso} Security Tools}
In 2013, Bai et al.~\cite{authscan} have proposed AuthScan, a framework to extract the authentication protocol specifications automatically from implementations.
They found security flaws in several \gls{sso} systems.
The authors concentrated on \gls{mitm} attacks, Replay attacks and Guessable tokens.
More complex attacks, like \gls{ids} or \gls{kc}, cannot be evaluated.
In the same year, Wang et al.~\cite{integuard} developed a tool named \emph{InteGuard} detecting the invariance in the communication between the client and \gls{sp} to prevent logical flaws in the latter one.
Another tool similar to \emph{InteGuard} is \emph{BLOCK}~\cite{block_brm}, which acts as a proxy and examines to the invariance of web related messages.
Both tools should be able to detect Replay attacks and \gls{trc}.
Since all HTTP messages between the adversary and the \gls{sp} are valid and do not show abnormalities, neither \emph{InteGuard} nor \emph{BLOCK} is able to mitigate \gls{ids}, \gls{ds}, \gls{kc} and XML External Entity.
Evans et al.~\cite{ssoscan} published on USENIX'14 a fully automated tool named \emph{SSOScan} for analyzing the security of \gls{oauth} implementations and described five attacks, which can be automatically tested by the tool.

\section{Lessons Learned}
\label{sec:lessons_learned}

\begin{overview}
Now, despite the above concern, to me, it is clear that the issues you bring up are still important and we can learn a lot from them. 
But, the paper doesn't help me learn from these issues, instead it only tells me all the issues. 
When I read the abstract, I was excited particularly by the "better understand implementation issues in \gls{oid} applicable to other  systems" and "we propose generic fixes". 
On reading this in the abstract, I was hoping a discussion on how the issues you noticed apply to other protocols and \glspl{sso}, or what we can learn as a community. 
Further, I was also hoping for a discussion on mitigations in both current implementations and future libraries and proposals.

I was disappointed to find this discussion missing from the paper. 
I feel like these discussions would significantly strengthen the paper and make it a great contribution fitting in a top-tier conference. 
For example, the flaws you found seem to be implementation flaws or engineering flaws in cryptographic protocols. 
Have you considered connecting the flaws to the classic "Prudent Engineering Practices for Crypto Protocols" paper from 1995? Are there cases where the \gls{oid} specification itself ignored some of the recommendations?

The paper I mention above is just an example, but I would have loved to see principles or anti-patterns  extracted from all the flaws you found. 
Additionally, if you could track down anti-patterns in the spec (vs. coding anti-patterns) that would be even more awesome. 
Can we also learn lessons for \gls{sso} protocols like \gls{oauth}? What about \gls{oid} 2.0 (based on \gls{oauth})? It seems like there is a lot of work on broken web-based \glspl{sso} and we really need to find out what mistakes we are making in designing them. 

For example, one anti-pattern might be that asking the \gls{sp} to verify multiple things in the token instead of just checking a simple HMAC might reduce the bugs noticed TRC attacks. 
Or, can we add something to browser APIs to improve this current state of affairs? Or, the KC attack Strategy 1, is clearly because of bad design: the table at the \gls{sp} shouldn't be keyed by \alpha. 
More discussions like these would make a great contribution.
\end{overview}

\paragraph{Trusted \glspl{idp}}
When Microsoft introduced MS Passport, the first web \gls{sso} system, criticism concentrated on the closed nature of the system: only a single \gls{idp} at the domain {\tt passport.com} was used. Thus subsequent approaches like MS Cardspace and \gls{saml} Web \gls{sso} allowed multiple \glspl{idp}, but still retained the idea that an \gls{idp} should only be run by trusted parties, and that a trust relationship between a \gls{sp} and an \gls{idp} should be established manually. With \gls{oid}, ``openness'' for the first time became more important than ``trustwothyness'', and this resulted in new attack classes. The lesson learned is that the establishment of trust should not be fully automated, if this isn't backed up by solid cryptography (like e.g. in PKI scenarios).

\paragraph{Identities are Important}
Attacks similar to \gls{trc} have been described before in the literature. E.g. Armado et al. discovered a bug in the Google \gls{sso} implementation where the identity of the target \gls{sp} was omitted from the \gls{saml} assertion. Thus an assertion issued for (low-security) service A (controlled by the attacker) could be used to log into (high-security) service B.
Including identities in protocol messages, and checking these values, is good engineering practice, e.g. in TLS certificate verification. The lesson learned from the \gls{trc} attack is that checking identity of the \gls{sp} is always important and should be enforced in any application.

\paragraph{References to Cryptographic Keys}
\gls{kc} exploits weaknesses in the association between the identity of the \gls{idp}, the key handle and the key value used for the signature verification. 
In \gls{oid} the only connection between the key and the corresponding \gls{idp} is the \gls{asso} handle $\alpha$. 
Unfortunately, the value of $\alpha$ can be freely chosen by \emph{any} \gls{idp}.
In case of \gls{oid}, if the loading of the key occurs only on basis of $\alpha$ and without verifying the corresponding \gls{idp}, \gls{kc} is applicable. 
Lessons learned: The identification of the correct cryptographic keys should be unambiguous. If keys are related to the identity of a communicating party, then this identity should be part of the key identifier. E.g., keys should be stored indexed by a pair $(IdP_{ID},\alpha)$.

\paragraph{Multiple Equivalent Parameters}
If two or more different parameters are used for the same purpose, then it is difficult to formally specify how to react if these two parameters have different semantics. This fact was exploited in the Discovery Spoofing attack, which is only possible if two different strings are used as identifiers for the same entity. Similar problems have been reported in multi-layer messaging: E.g. in SOAPAction Spoofing, the SOAP action can be specified in the HTTP and in the SOAP Header. By specifying two different values, inconsistent behaviour from the SOAP receiver can be triggered.

\paragraph{Complex Information Flow Specification}
In many cases, developers of \gls{oid} frameworks deviated from the specification, which resulted in a different, vulnerable message flow. It seems that the \gls{oid} specification is not clear enough to unambiguously implement the desired message flow. It is an interesting open question how to formally specify the desired flow, such that computer-aided enforcement of this flow, or computer-aided checking of this flow, becomes possible.

\section{Future Work}
\label{sec:conclusion}

\begin{overview}
    \begin{itemize}
        \item In this paper, we present a new adversarial model for SSO
        \item In this model, the attacker controls a custom IdP as well as a client
        \item Each discovered flaws allows an attacker to log in as the victim.
        \item We applied our model to the OpenID protocol and created the open source tool \emph{OpenID Attacker}, which acts as a malicious IdP.
        \item Our approach allowed to flexible test different OpenID SPs, including frameworks, libraries and websites.
        \item We showed that the majority of available frameworks and libraries are vulnerable.
        \item Additionally, we revealed new classes of OpenID attacks.
        \item We reported our findings to the developers and they acknowledged its the importance.
        \item In future work, we plan to adopt the adversarial model to different SSO protocols like Facebook connect, OAuth, SAML, BrowserID, BrowserID, \dots
    \end{itemize}
\end{overview}

We showed that \gls{sso} protocols and implementations are a high-value attack target.
Although there is a lot of research in the area of \gls{sso}~\cite{sun2012devil,somorovskySAML,ssoscan} and \gls{oid}~\cite{microsoft,journals/compsec/SunHB12,kohlarOpenid}, the number of  vulnerabilities found is surprisingly high. 

We believe that the concept of a malicious \gls{idp} is a threat to all open \gls{sso} protocols, thus future work includes applying the methodology developed in this paper to different protocols like OAuth, SAML an OpenID Connect.

%In future work, we suggest 
%\begin{inparaenum}
% \item to start a large scale study on more online \gls{sp} deployments by using the \gls{oida}, and
% \item to adopt our \gls{sso} attacker paradigm to different \gls{sso} protocols like \gls{saml}, \gls{oauth}, OpenIDConnect and BrowserID.
% % \item Based on the knowledge we gained on critical steps in \gls{sso} protocols, we want to make first steps in developing  SSOAttacker -- a tool for automated analysis of several \gls{sso} protocols, covering all related attacks.
%\end{inparaenum}

We will make the source code of \gls{oida} public, encouraging researchers and penetration tester to use this tool to further improve security in \gls{sso} systems, and to adapt it to other protocols.

%ACKNOWLEDGMENTS are optional
\section{Acknowledgments}
The authors would like to thank Juraj Somorovsky for fruitfull discussions and the motivation to go on further and deep on this topic.
We additionally want to thank Christian Kossmann for the extension of \gls{oida} to support fully automatic \gls{oid} attacks.

\bibliographystyle{IEEEtran}
\bibliography{references,nds_publications/nds_publications_2008,nds_publications/nds_publications_2010,nds_publications/nds_publications_2011,nds_publications/nds_publications_2012,nds_publications/nds_publications_2013,nds_publications/nds_publications_2014,bib/literature,bib/rfc}
% ACM needs 'a single self-contained file'!
%\balancecolumns
\newpage
\appendix
\begin{table}[h!]
    \centering
\begin{tabular}{|l|p{0.68\linewidth}|}
\hline
Notation & Explanation \\
\hline
$\URLID$ & A URL representing a user's \emph{login name} \\
$\URLID_\C$ & A URL representing $\C$'s \emph{login name} at $\URLIdP_\C$ \\
$\URLID_\A$ & A URL representing $\A$'s \emph{login name} at $\URLIdP_\A$ \\
$\URLSP$ & The URL of the \gls{sp}, e.g.\ \url{http://mysp.com} \\
$\URLSP_\A$ & The URL of the attacker $\A$ controlled \gls{sp}, e.g.\ \url{http://sp.attacker.com} \\
$\URLIdP$ & The URL of the user's \gls{idp}, e.g.\ \url{https://www.google.com/accounts/o8/ud} \\
$\URLIdP_\C$ & The URL of $\C$'s \gls{idp}, e.g.\ \url{https://www.google.com/accounts/o8/ud}. \\
$\URLIdP_\A$ & The URL of the attacker $\A$ controlled \gls{idp}, e.g.\ \url{http://idp.attack.com}. \\
$t$ & The \gls{oid} token, containing at least $\URLID$, $\URLSP$ and $\URLIdP$. \\
$\sigma$ & The signature value for token $t$. \\
$\alpha$ & The value $\alpha$ is used to identify the key to verify $(t,\sigma)$. Note that $\alpha$ is just a reference value to the key and does not contain any key material. 
For the attack on Drupal, we also used $\beta$, because there are two different \glspl{asso}.\\
\hline
\end{tabular}
\vspace{2mm}
\caption{List of notations used in this paper.}
\label{tab:notations}
\end{table}

% \begin{table}[h!]
    % \centering
    % \begin{tabular}{|l|p{0.7\linewidth}|l|}
        % \hline
        % Protocol/Framework & URL Redirect Parameter & Source \\
        % \hline
        % \gls{saml} & \\
        % \gls{oid} & \texttt{openid.return\_to} ($\URLSP$ in this paper) \cite[Section 9.1]{openid20}\\
        % \gls{oauth} & \texttt{redirect\_uri} & \cite[Section 4.1.3]{rfc6749oauth} \\
        % \hline
    % \end{tabular}
    % \caption{List of parameters that contain \emph{routing information} categorized by \gls{sso} protocol/framework.}
    % \label{tab:redirect_parameters}
% \end{table}

\end{document}